\documentclass[twocolumn,
superscriptaddress,
amssymb,
nobibnotes,
aps,
showkeys,
longbibliography,
prd]{revtex4-2}

\usepackage{amsmath}
\usepackage{amsfonts}
\usepackage{amssymb}
\usepackage{amsthm}
\usepackage{booktabs}
 \usepackage[unicode=true,pdfusetitle,
  bookmarks=true,bookmarksnumbered=false,bookmarksopen=false,
  breaklinks=false,pdfborder={0 0 1},backref=false,colorlinks=true]
  {hyperref}
\usepackage[dvipsnames,table]{xcolor}
\usepackage{color}
\hypersetup{
  colorlinks=true,
  urlcolor=NavyBlue,
  citecolor=Orchid,
  linkcolor=Magenta}
\usepackage{graphicx}
\usepackage{subfigure}
\usepackage{rotating}
\usepackage{xcolor}
\usepackage{tabularx}
\usepackage{multirow}

\usepackage{circuitikz}
\usetikzlibrary{positioning,fit,calc}

\tikzstyle{block} = [draw=black, thick, minimum width=1.618cm,
minimum height=1cm, align=center, rounded corners=3]

\tikzstyle{line} = [-,>=stealth]

\usepackage{adjustbox}

\newcommand{\geant}{G\textsc{eant4}~}
\setcitestyle{maxnames=3}

\begin{document}

\title{Germanium Atomic Compton Scattering Measurements and \emph{ab
    initio} Many-Body Calculations: Implications for Electronic recoil Dark Matter
  Detection}%

\author{Chang-Hao~Fang}%
\affiliation{College of Physics, Sichuan University, Chengdu 610065}

\author{Yi-Ke~Shu}
\affiliation{College of Physics, Sichuan University, Chengdu 610065}

\author{Shin-Ted~Lin}
\email[Corresponding Author: ]{stlin@scu.edu.cn}
\affiliation{College of Physics, Sichuan University, Chengdu 610065}

\author{Shu-Kui~Liu}
\email[Corresponding Author: ]{liusk@scu.edu.cn}
\affiliation{College of Physics, Sichuan University, Chengdu 610065}

\author{Hao-Yang~Xing}
\affiliation{College of Physics, Sichuan University, Chengdu 610065}

\author{Jing-Jun~Zhu}
\affiliation{College of Physics, Sichuan University, Chengdu 610065}

\author{Hsin-Chang~Chi}
\affiliation{College of Physics, National Dong Hwa University, Shoufeng 97401}

\author{Muhammed~Deniz}
\affiliation{Department of Physics, Dokuz Eyl\"ul University, 35160 Buca,
  \.Izmir}

\author{Hai-Tao~Jia}
\affiliation{College of Physics, Sichuan University, Chengdu 610065}

\author{Han-Yu~Li}
\affiliation{College of Physics, Sichuan University, Chengdu 610065}

\author{Qian-Yun~Li}
\affiliation{College of Physics, Sichuan University, Chengdu 610065}

\author{Ren-Ming-Jie~Li}
\affiliation{College of Physics, Sichuan University, Chengdu 610065}

\author{Yu~Liu}
\affiliation{College of Physics, Sichuan University, Chengdu 610065}

\author{Xiao-Yu~Peng}
\affiliation{College of Physics, Sichuan University, Chengdu 610065}

\author{Hao-Yu~Shi}
\affiliation{College of Physics, Sichuan University, Chengdu 610065}

\author{Qin~Wang}
\affiliation{College of Physics, Sichuan University, Chengdu 610065}

\author{Henry~Tsz-King~Wong}
\affiliation{Institute of Physics, Academia Sinica, 11529 Taipei}

\author{Yu-Lu~Yan}
\affiliation{College of Physics, Sichuan University, Chengdu 610065}

\author{Li-Tao~Yang}
\affiliation{Key Laboratory of Particle and Radiation Imaging (Ministry of
  Education) and Department of Engineering Physics, Tsinghua University,
  100084 Beijing}

\author{Qian~Yue}
\affiliation{Key Laboratory of Particle and Radiation Imaging (Ministry of
  Education) and Department of Engineering Physics, Tsinghua University,
  100084 Beijing}

\date{\today}%

\begin{abstract}
  Diverse searches for direct dark matter (DM) in effective
  electromagnetic and leptophilic interactions resulting from new physics,
  as well as Weakly Interacting Massive Particles (WIMPs) with
  unconventional electronic recoils, are intensively pursued.
  Low-energy backgrounds from radioactive $\gamma$ rays via Compton
  scattering and photon coherent scattering are unavoidable in terrestrial
  detectors.
  The interpretation of dark matter experimental data is dependent on a
  better knowledge of the background in the low-energy region.
  We provide a 2.3\% measurement of atomic Compton scattering in the low
  momentum transfer range of 180 eV/c to 25 keV/c, using a 10-g
  germanium detector bombarded by a $^{137}\mathrm{Cs}$ source with a
  7.2 m-Curie radioactivity and the scatter photon collected by a
  cylindrical NaI[Tl] detector.
  The ability to detect Compton scattering's doubly differential cross
  section (DDCS) gives a special test for clearly identifying the
  kinematic restraints in atomic many-body systems, notably the Livermore
  model.
  Additionally, a low-energy-background comparison is made between
  coherent photon scattering and Compton scattering replacing the
  scattering function of \geant\@software, which uses a completely
  relativistic impulse approximation (RIA) together with
  Multi-Configuration Dirac-Fock (MCDF) wavefunctions.
  For the purpose of investigating sub-GeV mass and electronic-recoil dark
  matter theories, signatures including low energy backgrounds via high
  energy $\gamma$ rays in germanium targets are discussed.
\end{abstract}

\maketitle

\section{Introduction}
\label{sec:intro}

Weakly interactive massive particles (WIMPs) coupling with atomic nuclei
($\chi$-N) has garnered experimental interest over the past few
decades~\cite{goodmanDetectabilityCertainDarkmatter1985}.
Recent null results from searches at high
mass ranges~\cite{jiangLimitsLightWeakly2018, aprileFirstDarkMatter2023,
  aalbersFirstDarkMatter2023, mengDarkMatterSearch2021}, as well as
techniques sensitive to single electron-hole
pairs~\cite{tiffenbergSingleElectronSinglePhotonSensitivity2017,
  renDesignCharacterizationPhononmediated2021},
have generated significant interest within the community regarding light
dark matter coupling with electrons
($\chi$-e)~\cite{arnquistFirstConstraintsDAMICM2023,
  collaborationSENSEIFirstDirectDetection2023,
  arnaudFirstGermaniumBasedConstraints2020}.
Underground facilities offer muon-free experimental
environments~\cite{chengChinaJinpingUnderground2017}, and
additional specialized auxiliary facilities are being utilized to further
minimize background level.
However, long-lived U/Th decay chain isotopes from rocks and materials
surrounding detectors continuously emit $\gamma$ rays, which contribute to
near-threshold background through Compton scattering (CS) and photon coherency
scattering (PCS).
While certain techniques can differentiate between nuclear recoil and
electron recoil signals, this discrimination capability diminishes in the
near-threshold region.
Consequently, backgrounds from CS and PCS must be
considered equally.
An accurate assessment of the $\gamma$ background is crucial for establishing
constraints on low-mass dark matter.

The $\gamma$-induced structured background dominates the region of interest for
LDM through electronic final states.
The scattering dynamics of $\chi$-e interactions are, in principle,
indistinguishable from CS.
Unfortunately, this similarity results in both the Compton background and
the expected LDM energy spectra exhibiting analogous step-like structures
at the atomic ionization energies~\cite{barkerLowEnergyBackground2016,
  huMeasurementComptonScattering2023,
  norciniPrecisionMeasurementCompton2022}.
For energies below 200 eV, the cross-section of PCS increases rapidly,
making it a leading background~\cite{robinsonCoherentPhotonScattering2017,
  duSourcesLowEnergyEvents2022,
  berghausPhononBackgroundGamma2022}.
Furthermore, the kinematic cutoff of PCS introduces significant step-like
structures into the energy spectrum.
These similar structures complicate the identification of signals from
backgrounds.

Atomic CS involves the binding effects of electrons,
electron correlation, and quantum many-body effects.
The most common approach is the relativistic impulse approximation (RIA)
formalism~\cite{ribberforsRelationshipRelativisticCompton1975,
  ribberforsRelationshipRelativisticCompton1975a,
  ribberforsIncoherentxrayscatteringFunctionsCross1982}.
The RIA accurately describes the Doppler broadening and atomic binding
effects in CS~\cite{kleinUeberStreuungStrahlung1929,
  bergstromOverviewTheoriesUsed1997,
  brownLowEnergyBound2014}, while remaining simple to implement in Monte
Carlo (MC) algorithms~\cite{cullenSimpleModelPhoton1995,
  brownLowEnergyBound2014,nelsonEgs4CodeSystem1985,
  brusaFastSamplingAlgorithm1996}, such as
\geant~\cite{agostinelliGeant4SimulationToolkit2003}.
Current dark matter experiments utilize MC techniques to construct the
Compton background.
Therefore, the reliability of the RIA theory and the corresponding MC
models should be carefully scrutinized in the low-momentum transfer region.
Recent measurements of the Compton energy spectrum using semiconductor
detectors~\cite{norciniPrecisionMeasurementCompton2022,
  huMeasurementComptonScattering2023}, coupled with new theoretical
perspective~\cite{essigLowenergyComptonScattering2024} that account for
condensed matter effects, have provided valuable insight.
However, a comprehensive examination of the relevant theoretical models
from the perspective of the double-differential cross-section (DDCS)
remains essential.

From theoretical perspective, we employed a fully relativistic ab initio
atomic many-body treatment, namely multi-configuration Dirac-Fock (MCDF)
method~\cite{desclauxMulticonfigurationRelativisticDIRACFOCK1975}, to
reassess the atomic CS.
Compared with the previous Hartree-Fock (HF) results that adopted in \geant~\cite{biggsHartreeFockComptonProfiles1975,
  hubbellAtomicFormFactors1975,
  cullenEPDL97EvaluatedPhoto1997}, our results presents significant
differences in differential cross-section $d\sigma/d\Omega$ (DCS) for the small-angle
scattering~\cite{jiaHighaccuracyMeasurementCompton2022}.
In the context, we denote the RIA calculation with MCDF or HF atomic input
as ``MCDF-RIA'' and ``HF-RIA'', respectively.

Precise measurements were conducted to resolve ambiguities in low-energy
scenarios.
With precise control of the scattering angle ($<0.03^\circ$), we measured the
DDCS spectra across a range of angles from 1.5 to 12 degrees.
Measurement at the scattering angle of $1.5^\circ$ enabled us to investigate
CS at a momentum transfer of at least 180 keV/c.
A meticulous analysis of the background, efficiency, and systematic
uncertainties was performed.
Through these measurements, the low-energy Compton models implemented in
\geant and the inconsistencies in DCS were directly and thoroughly tested.

This paper is organized as follows.
In Section~\ref{sec:formalism}, we review the RIA, low-energy Compton
models, and photon coherent scattering calculations.
Section~\ref{sec:experimental-setup} describe our experimental setup.
Section~\ref{sec:data-analysis} presents our data analysis methods.
The experimental results compared with simulation data for DDCS and
measured incoherent scattering function are reported in
Section~\ref{sec:results-discussion}.
In the Section~\ref{sec:compt-scatt-DM-bkg}, we present a detailed
discussion of the impact of detector masses, different SFs, and $\gamma$-ray
positions on Compton backgrounds, as well as a combined analysis of
$\gamma$-induced electronic final-state backgrounds for various LDM models.

\section{RIA, Monte Carlo Models and Photon Coherent Scattering}
\label{sec:formalism}

\subsection{Relativistic Impulse Approximation for Compton Scattering}

\label{subsec:MCDF-RIA}

In the RIA formalism, the incoherent DCS
can be factorized as
follows~\cite{ribberforsRelationshipRelativisticCompton1975,
  ribberforsRelationshipRelativisticCompton1975a,
  ribberforsIncoherentxrayscatteringFunctionsCross1982,
  ribberforsXrayIncoherentScattering1983}

\begin{equation}
  \label{eq:RIA-dcs}
  \left[\frac{d\sigma}{d\Omega}\right]_\text{RIA} \simeq {\left[\frac{d\sigma}{d\Omega}\right]}_\mathrm{FEA}\cdot S(X),
\end{equation}
where the first term is the DCS under the free electron approximation
(FEA) and the subsequent term denotes the incoherent scattering function
(also referred to as the incoherent scattering factor).
The incoherent scattering function serves as a correction factor that
arises from the atomic system.
The variable $X$ is proportional to the momentum transfer, which is defined
by the incident photon wavelength $\lambda$ and the scattering angle
$\theta$~\cite{hubbellAtomicFormFactors1975,kahaneRelativisticDiracHartree1998}
\begin{equation}
  \label{eq:sf-X}
    X \equiv \frac{\sin \left(\theta/2\right)}{\lambda}.
\end{equation}
The scattering function can be obtained through the number of
electrons ($Z_i$), binding energy ($B_i$), and Compton profile ($J_i$) of
the $i$th sub-shell~\cite{hubbellAtomicFormFactors1975,
  blochAtomicShellCompton1974}
\begin{equation}
  \label{eq:scattering-function}
  S(X) = \sum_i Z_i \Theta\left(E-B_i\right) \int_{-\infty}^{p_i^{\max}} J_i(p_z)dp_z.
\end{equation}
The influence of the atomic system on the DCS primarily arises from two
aspects: the binding of electrons and the momentum structure of electrons.
The Heaviside function $\Theta$ ensures the atomic binding effect in the
scattering process.
The Compton profile, which characterizes the momentum distribution of
electrons, introduces the kinetic information prior to the
scattering~\cite{ribberforsIncoherentxrayscatteringFunctionsCross1982}.
Due to the spherical symmetry of the momentum distribution, only the
$z$ direction (the direction of photon incidence) is commonly considered.

The DCS of RIA closely depends on the atomic ground-state.
To achieve a more accurate scattering behavior, we conducted fully
relativistic ab-initio atomic many-body MCDF calculations for Ge.
The MCDF method allows the electron correlations and configuration
interactions being sufficiently
considered~\cite{desclauxMulticonfigurationRelativisticDIRACFOCK1975}.

We reassess the binding energies, Compton profiles, and incoherent
scattering function using MCDF wavefunctions.
In the remainder of the section, we will discuss these comparisons and
their their implications for RIA cross sections.
The ionization energies of each sub-shell from MCDF calculation, the EPDL
library, and experimental measurements are presented in
Table~\ref{tab:ionization-energy}~\cite{cullenEPDL97EvaluatedPhoto1997,
  henkeXRayInteractionsPhotoabsorption1993,
  deslattesXrayTransitionEnergies2003}.
A better agreement, particularly for outer shell electrons, between the
MCDF results and experimental data indicates the validity of incorporating
electron correlations and configuration interactions in the Ge system.
The discrepancies between the Compton profile obtained through the MCDF-RIA
and the Waller-Hartree formalism with the Hartree-Fock
wavefunction~\cite{biggsHartreeFockComptonProfiles1975} are thoroughly
investigated.
No significant differences were observed; however, some asymmetric
deviations were noted in the high-momentum region due to relativistic
effects.
The impact of these deviations is limited, as the Compton profile is three
orders of magnitude lower than that of the peak region.

\begin{table*}
  \caption{The ionization energies of sub-shells for Ge are
    presented, encompassing MCDF ionization energies, non-relativistic HF
    ionization energies, and experimentally-measured ionization energies.
    The MCDF ionization energies have been computed via our calculations,
    while the non-relativistic HF ionization energies have been obtained
    from the Evaluated Photo Data Library (EPDL)
    database~\cite{cullenEPDL97EvaluatedPhoto1997}.
    The experimental measurements have been sourced from studies conducted
    by \citet{henkeXRayInteractionsPhotoabsorption1993} and
    \citet{deslattesXrayTransitionEnergies2003}.
    Unit of these energies are eV.}\label{tab:ionization-energy}
  \begin{ruledtabular}
    \begin{tabular}{lcccccccccccc}
      Sub-shells & $K$ & $L_\mathrm{I}$& $L_\mathrm{IIa}$& $L_\mathrm{IIb}$ & $M_\mathrm{I}$ & $M_\mathrm{IIa}$ & $M_\mathrm{IIb}$ & $M_\mathrm{IIIa}$ & $M_\mathrm{IIIb}$ & $N_\mathrm{I}$ & $N_\mathrm{IIa}$ & $N_\mathrm{IIb}$ \\
               & $1s_{1/2}$ & $2s_{1/2}$ & $2p_{1/2}$ & $2p_{3/2}$ & $3s_{1/2}$ & $3p_{1/2}$ & $3p_{3/2}$ & $3d_{3/2}$ & $3d_{5/2}$ & $4s_{1/2}$ & $4p_{1/2}$ & $4p_{3/2}$ \\
      \cmidrule{1-13} 
      MCDF & 11119.0 & 1426.9 & 1257.3 & 1226.0 & 193.3 & 136.8 & 132.2 & 36.5 & 35.9 & 14.6 & 7.8 & 8.0 \\
      HF (EPDL) & 11067.0 & 1402.3 & 1255.4 & --- & 179.25 & 129.38 & --- & 38.19 & --- & 14.7 & 6.5 & --- \\
      Exp. & 11103.1 & 1414.6 & 1248.1 & 1217.0 & 180.1 & 124.9 & 120.8 & 29.9 & 29.3 &  --- & 7.9 & --- \\

    \end{tabular}
  \end{ruledtabular}
\end{table*}

Significant difference between the HF-RIA and MCDF-RIA methods is
identified in the scattering function for the low-momentum transfer (small
$X$) scenario.
Fig.~\ref{fig:scattering-fucntion} illustrates the scattering functions
for the two methods.
Theoretical analysis indicates that the scattering function is highly
sensitive to different binding energy, while it exhibits minimal
sensitivity to deviation in the Compton profile.
This difference in scattering function ($\mathrm{i.e.}$ DCS) would reveal a
new Compton background level in sub-keV regions.

\subsection{Low-energy Compton Models in \geant}
\label{subsec:g4-models}

\begin{figure}[]
  \centering
  \includegraphics[width=\linewidth]{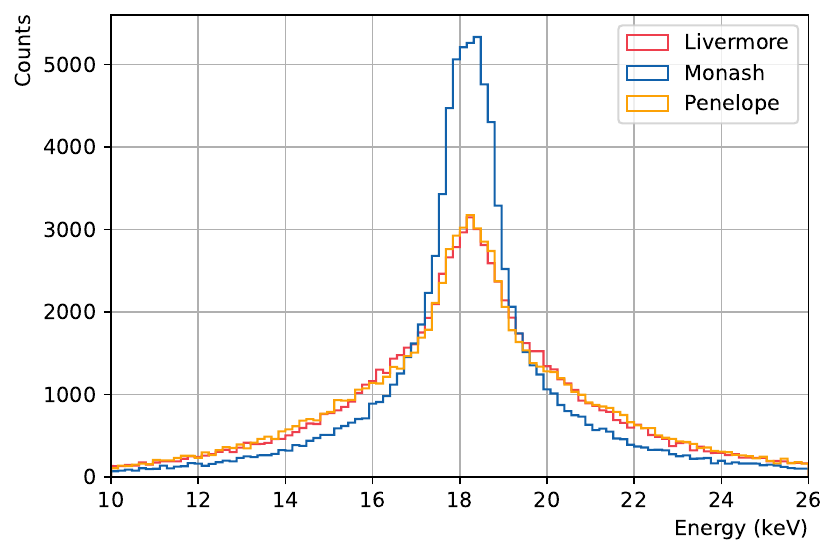}
  \caption{The DDCS sampled by three low-energy Compton
    scattering models implemented in \textsc{Geant4}.
    LivermoreComptonModel (red), MonashComptonModel (bule) and
    PenelopeModel (orange) at the scattering angle of ${(12\pm0.2)}^\circ$.}
  \label{fig:model-comparason}
\end{figure}

Three low-energy Compton models, namely the LivermoreComptonModel
(Livermore model~\cite{apostolakisGEANT4LowEnergy1999}),
LowEPComptonModel (Monash model~\cite{brownLowEnergyBound2014}), and
PenelopeModel (Penelope model~\cite{salvatgavaldaPENELOPE2008Code2009})
have been implemented in \geant
(version 10.05~\footnote{And still have three models in version (11.03)})
based on the RIA~\cite{agostinelliGeant4SimulationToolkit2003}.
A significant difference between the Monash model and the other two models
is that the Monash model does not require the directions of the outgoing
photon and electron to remain in the incident plane.
As illustrated in Fig.~\ref{fig:model-comparason}, this choice results in
the DDCS, being more
concentrated around the Compton peak in the low-energy region, since only
the projection of the electron momentum participates in the scattering
process~\cite{brownLowEnergyBound2014,salvatgavaldaPENELOPE2008Code2009}.
The evaluation of its impact on the energy spectra (DCS, $d\sigma/dT$) revealed
no significant differences among the three models.
However, for angle-dependent simulations, such as those involving
anti-coincidence systems, potential differences warrant further
attention.
The discrepancies in DDCS among the three models in the low-energy regime
will be experimentally inspected in Section~\ref{sec:ddcs}.

In this work, the binding energies and scattering functions in \geant have
been replaced with experimental values and MCDF-RIA results, respectively.
Given that the Livermore and Penelope models demonstrate no significant
differences regarding the issues of interest in this study, subsequent
analyses will focus exclusively on the Livermore model.

\subsection{Photon Coherent Scattering}
\label{sec:phot-coher-scatt}
PCS describes the phenomenon in which the
scattered particle maintains a fixed phase relationship with the initial
state.
In the region of interest for LDM detection, the cross-section for PCS
increases in the sub-keV range~\cite{robinsonCoherentPhotonScattering2017,
  robinsonErratumCoherentPhoton2017}.
The cross-sections of PCS are coherently combined with
component amplitudes~\cite{royElasticScatteringPhotons1999}.
Rayleigh scattering, nuclear Thomson scattering, and Delbr{\"u}ck
scattering, representing scattering by bound electrons,
nuclei, and positronium created through vacuum polarization, participate as
components of the PCS~\cite{kaneElasticScatteringGrays1986}.

Rayleigh scattering, the dominance of PCS, is obtained by Thomson
scattering cross-section with correction from atomic form
factor~\cite{kaneElasticScatteringGrays1986,
  schauppSmallAngleRayleigh1983},
\begin{equation}
  \label{eq:Rayleigh-DCS}
  {\left[\frac{d\sigma}{d\Omega}\right]}_\mathrm{Rayl.} = \frac{1}{2}r_e^2(1+\cos^2\theta)\cdot\left|f(q, Z)\right|^2.
\end{equation}
where $r_e$ is classical electron radius, $\theta$ is scattering angle, and $f(q, Z)$ is the relativistic atomic form factor.
It can be expressed as
\begin{equation}
  \label{eq:Rayl-form-factor}
  f(q,Z) = 4\pi\int_0^\infty \rho(r,Z) \frac{\sin(qr)}{qr}r^2dr,
\end{equation}
in which $\rho(r,Z)$ represents the charge distribution and $q$ represents
photon momentum transfer.
The Thomson nuclear scattering can be
expressed as~\cite{kaneElasticScatteringGrays1986}
\begin{equation}
  \label{eq:Thomson-DCS}
  \left[\frac{d \sigma}{d \Omega}\right]_{\mathrm{Thom.}}=\frac{1}{2} r_e^2\left(1+\cos ^2 \theta\right)\left(\frac{m}{M} f_{\mathrm{nuc}}\right)^2,
\end{equation}
where $m$, $M$ represent mass of electron, nuclear respectively,
and $f_{\mathrm{nuc}}$ is nuclear form factor. In this work, we adopt $Z^2$
nuclear form factor under the point charge approximation.

The DCS of Delbr{\"u}ck scattering can be written as

\begin{equation}
  \label{eq:delbruck-DCS}
  \left[\frac{d \sigma}{d \Omega}\right]_{\mathrm{Delb.}}=(\alpha Z)^2 r_0^2|a|^2.
\end{equation}
In this work, the DCSs are obtained via applying linear
interpolation on tabulated data from Ref.~\cite{falkenbergAmplitudesDelbruckScattering1992}, where the lowest Born-approximation is adopted.
In addition, for energies above the K edge, the DCS is a smooth
function of photon energy, atomic number, and scattering angle.
Cubic-spline is used to interpolate the cross-section
as a function of photon energy, using data from table in
Ref.~\cite{chatterjeeTablesElasticScattering1998}.
\begin{figure}[tb]
  \centering
  \includegraphics[width=\linewidth]{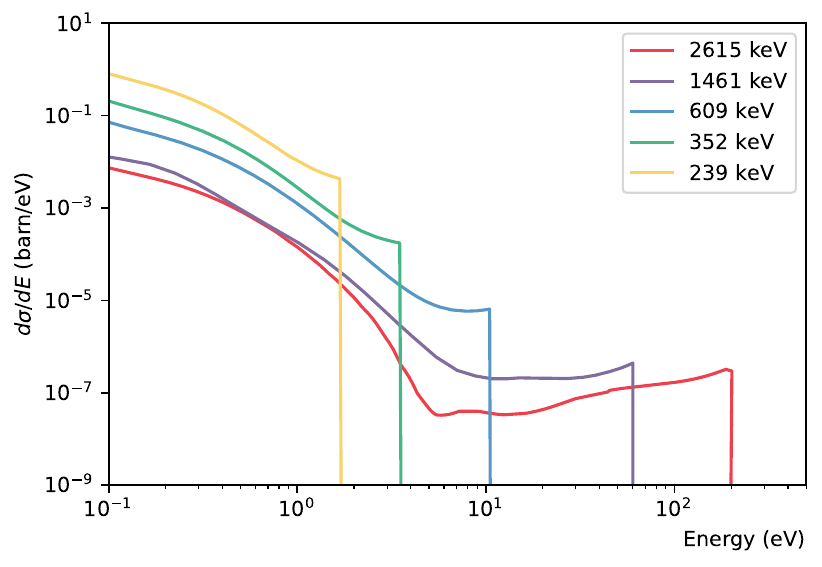}
  \caption{The differential cross-sections of photon coherent scattering
    with incident energy of 239 keV, 352 keV, 609 keV, 1461 keV and 2615
    keV. The sharp edge at the end of the DCS is from kinematic constrain.}
  \label{fig:dcs-phot-coher-scatt}
\end{figure}

The total cross-section of PCS should be obtained from the relativistic
second-order S-matrix calculations.
However, attribute to computational expensiveness of S-matrix methods,
particularly in the condition of high incident photon energies and atomic
many-body systems, other approaches have been explored to obtain scattering
cross-sections.
Form factors, which describe charge distributions, dispersion relations,
and the optical theorem that relates anomalous scattering amplitudes to the
total photon-atom cross-sections, have also been
explored~\cite{royElasticScatteringPhotons1999}.
However, the results obtained from form factor calculations often differ
significantly from those obtained using the S-matrix calculations,
particularly at large angles.

Contribution of three component amplitude depends on incidnent energy as
well as scattering angles.
In the energy range of 1 to 4 MeV, the amplitudes of different components
exhibit strong interference.
Below 1 MeV, Rayleigh amplitudes prevail at most scattering angles and
retain significant contribution at small angles with increasing incident
energy.
For relative large scattring angle, the contribution from nuclear Thomson
scattering is dominate and gradually becomes the primary factor at most
scattering angles as incident energy increasing.
The Delbr{\"u}ck amplitudes begin to contribute at intermediate angles below 1 MeV and, at higher energy, become important at intermediate and large
angles.
While Rayleigh and nuclear Thomson amplitudes generally interfere
constructively, the Delbrück amplitudes interfere destructively with
Rayleigh and nuclear Thomson amplitudes at small angles and constructively
at large angles.

Fig.~\ref{fig:dcs-phot-coher-scatt} illustrates the DCS of photon
coherent scattering.
The kinematic constrain of elastic scattering limits the maximum
momentum transfer, resulting a sharp edge in the end of spectra.
The total cross section increases as the photon energy decreases; however,
the energy range contributing to the background shifts as the photon energy
increases.
A combined analysis with Compton background is performed in
Section~\ref{sec:channels-bkgs}.

\section{Experimental setup}
\label{sec:experimental-setup}
To clarify the discrypancy between the CS functions and
models, a coincidence-based, high-accuracy experiment is designed and
performed.

\subsection{Experimental Apparatus}
\begin{figure*}[]
  \centering
  \includegraphics[width=\textwidth]{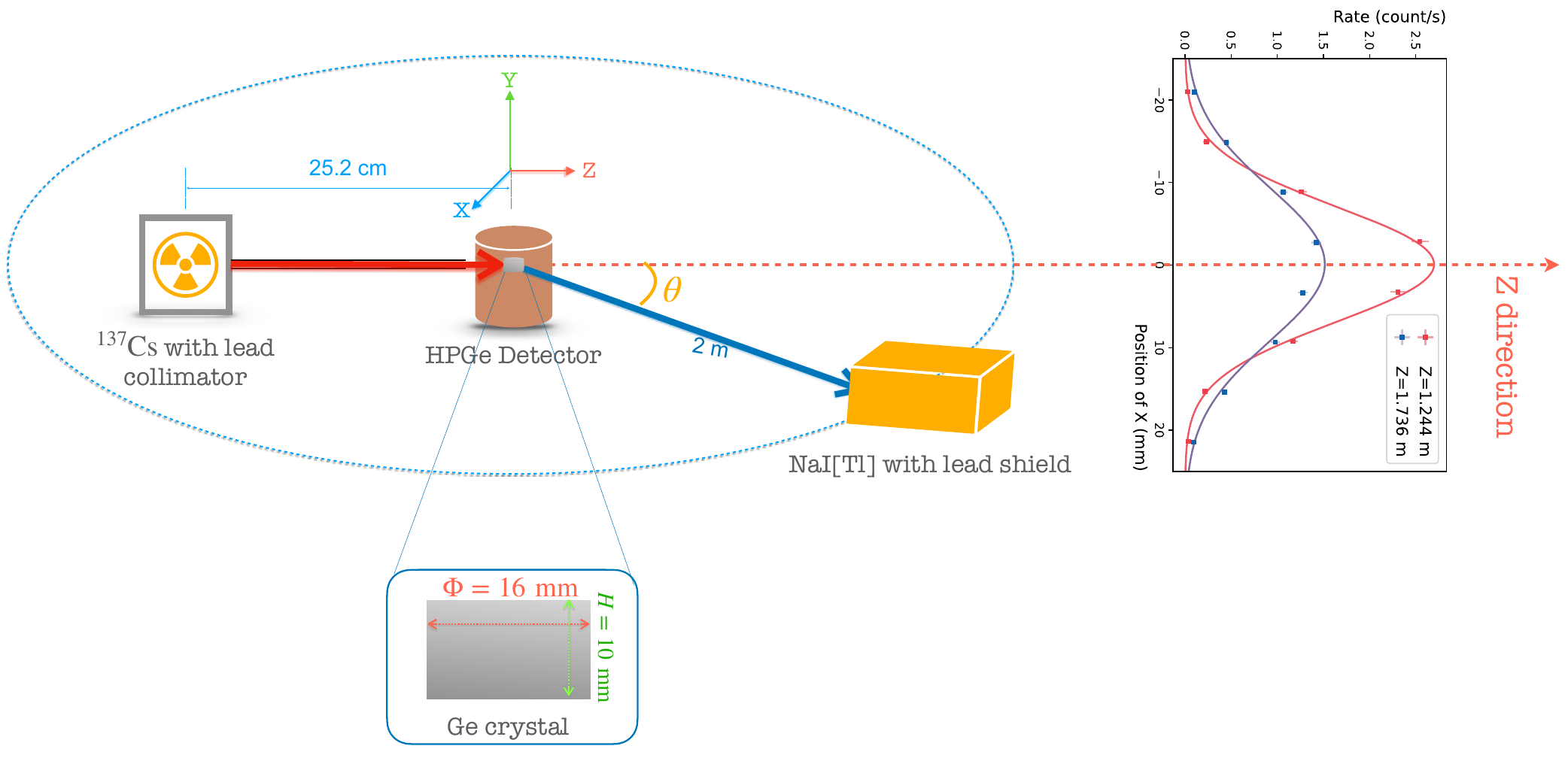}
  \caption{The schematic diagram of experimental design and apparatus.
    The detection system comprises a 7.18 mCi $^{137}\mathrm{Cs}$ source,
    an HPGe detector, a NaI[Tl] detector, and corresponding shieldings.
    All of the apparatus are situated on a calibrated horizontal
    experimental platform, denoted by the blue ellipse.
    The coordinate origin for this work is choosen at the center of the Ge
    crystal, with the $x$-axis lying along the platform.
    The z-axis is determined by two measurements illustrated in the right
    figure (details can be found in the text).}
  \label{fig:exp-schematic}
\end{figure*}

Fig.~\ref{fig:exp-schematic} illustrates the experimental apparatus for
measuring the CS spectrum at a specific scattering angle.
A collimated beam of 662 keV $\gamma$ rays from a $^{137}\mathrm{Cs}$
radioactive source with an activity of 7.18 mCi is used.
The $^{137}\mathrm{Cs}$ $\gamma$ source is embedded inside the Pb collimator,
whose collimating hole diameter is 3.8 mm.
The entire source apparatus is mounted on a lift table to fit the
experimental plane.

The detection system comprises a 10g n-type HPGe detector (16 mm diameter,
10 mm height) at the front end and a NaI[Tl] scintillator (76 mm diameter,
120 mm length) at the rear end.
The HPGe detector, surrounded by a cryostat made of oxygen-free high conductivity (OFHC) copper, is placed at the end of the stainless steel tube.
The rear-end NaI[Tl] detector is positioned 2 m from the HPGe detector
at a specific scattering angle.

\subsection{Scattering Angle Calibration}
\label{subsec:angle-clibration}

Accurate control of the scattering angle is a crucial aspect of our
approach, as measurements of DDCS and SF are susceptible to the scattering
angle.
To achieve a better of accuracy, we conducted a meticulous angle
calibration, the details of which are outlined below.

\textbf{Determine the horizontal plane.}
We used a laser level with a precision of $\pm0.3\ \mathrm{mm}/\mathrm{m}$
,corresponding to an angular precision of $\pm0.1^\circ$.
Once the horizontal plane was established, all detectors would be
meticulously positioned within this plane.

\textbf{Determine the experimental scattering angle of 0.}
The scattering angle of 0 is defined as the incident direction of photons
emitted by $^{137}\mathrm{Cs}$.
To determine this direction, we used a small cubic NaI detector
($6\times6\times6 \mathrm{mm}$) located in the horizontal plane calibrated
in the previous step.
The detector measured the count rate of 662 keV $\gamma$ while moving along
the $x$-axis at $Z=1.244\ \mathrm{m}$ and $1.736\ \mathrm{m}$, respectively
(see sub-figure in Fig.~\ref{fig:exp-schematic}).
The measured count rates were fitted with a Gaussian distribution to find
the maximum count rate which represents the center location of this
measurement.
The line of the zero scattering angle was defined by connecting these two
center locations.

\textbf{Calibrate the scattering angle $\theta$.}
The HPGe detector is fixed along 0-degree line at $Z=0$.
The experimental scattering angle is defined as the angle between the line connecting the HPGe and NaI[Tl] detectors and the 0-degree line.
Two laser levels were utilized to calibrate this scattering angle.
The first laser indicated the 0-degree line, while the second laser
indicated the scattering direction.
The two lasers converged at the center of the HPGe detector.
Once the direction was established, the NaI[Tl] detector was positioned as
far as possible to minimize systematic errors in the scattering angle,
which in this case was 2 m.

It is worth mentioning that the collimated $\gamma$ beam still experiences
some angle separation.
Through the measurements that determined the zero-scattering angle,
we quantified the deviation of the beam from the standard deviation
of the Gaussian fit, which was found to be $0.27\pm0.01^\circ$ at
$Z=1.244\ \mathrm{m}$ and $0.27\pm0.01^\circ$ at $1.736\ \mathrm{m}$,
respectively.
This separation results in a scattering angle that does not exactly match
the experimental calibration but follows a distribution.
To account for this source separation, we incorporated it into the
\geant\@simulation to obtain the actual effective scattering angle.

\subsection{Data Acquisition System}
Fig.~\ref{fig:DAQ} illustrates the electronics and data acquisition (DAQ)
system.
The signal is read out from the p+ contact by a low-noise field-effect
transistor (FET) located in the vicinity of Ge bulk, serving as the input
to reset the pre-amplifier.
The pre-amplifier has a single output that is distributed to the shaping
amplifier (S.A.).
In this experiment, we set the shaping time to
$2\ \mathrm{\mu s}$ ($\mathrm{SA}_2$) for all scattering angles except for the
measurement at $1.5^\circ$, which is set to $6\ \mathrm{\mu s}$ ($\mathrm{SA}_6$) to
achieve batter energy resolution.

Our S.A. outputs two signals.
One signal, passing through the leading edge discriminator, is fed into the
``AND'' logic unit, while the other signal is recorded by a 250 MHz Flash
Analog-to-Digital Converter (FADC) with 14-bit voltage resolution.
The recording time window is $40\ \mathrm{\mu s}$ for $\mathrm{SA}_2$ and
$80\ \mathrm{\mu s}$ for $\mathrm{SA}_6$.

Similarly, the NaI[Tl] detector also generates two signals.
One signal, passing through the leading edge discriminator, is fed into the
``AND'' logic unit, while the FADC records the other one.

A random trigger, generated by a pulse generator with a
frequency of 0.2 Hz, is recorded to derive the DAQ dead time and calibrate
the zero energy point.
This trigger is also utilized to estimate the data selection efficiency in
analyses.

\begin{figure}
  \centering
  \begin{adjustbox}{width=\linewidth}
    \begin{circuitikz}[transform shape]

      \node[block, fill=lightgray] (HPGe) {HPGe};
      \node[block, fill=lightgray, below=2cm of HPGe.center] (NaI) {\texttt{NaI[Tl]}};
      \node[block, fill=orange, above=2cm of HPGe.center] (RT) {\texttt{R.T.}};

      \node at ($(HPGe.center)+(2.5,+1)$) [block, fill=pink] (INH) {\texttt{Inh.}};
      \node at ($(HPGe.center)+(2.5,-1)$) [plain mono amp, scale=0.8, fill=red!30!white, thick] (A) {\texttt{S.A.}};
      \node at ($(RT.center)+(5, 0)$) [block, fill=lime] (Disc1) {\texttt{Disc.}};
      \node at ($(HPGe.center)+(5, 0)$) [block, fill=lime] (Disc2) {\texttt{Disc.}};
      \node at ($(Disc2.center)+(0, -2)$) [block, fill=lime] (Disc3) {\texttt{Disc.}};
      \node at ($(A.center)+(5, 0)$) [ieeestd and port, fill=cyan, fill opacity=0.7, thick] (AND) {\texttt{AND}};
      \node at ($(AND.out)+(0.375, 0)$) [ieeestd or port, fill=cyan, fill opacity=0.7, anchor=in 2, thick] (Or2) {\texttt{OR}};
      \node at ($(12.5, 0)$) [draw=black, thick, minimum width=1.618cm, minimum height=7cm, align=center, fill=orange!30!white] (FADC) {\texttt{FADC}};
      \node at ($(A.out)$) [circ] (corss-node) {};

      \draw [line, rounded corners=2, thick] (RT) -- (Disc1);
      \draw [line, rounded corners=2, thick] (HPGe.east) -- +(0.45, 0) |- (A.in);
      \draw [line, rounded corners=2, thick] (HPGe.east) -- +(0.45, 0) |- (INH.west);
      \draw [line, rounded corners=2, thick] (A.out) |- (Disc2.west);
      \draw [line, rounded corners=2, thick] (A.out) |- ($(FADC.west)+(0, -2.7)$) node[circ]{};
      \draw [line, rounded corners=2, thick] (Disc1) -- +(3.8,0) |- (Or2.in 1);
      \draw [line, rounded corners=2, thick, color=blue] (INH.east) -| (Or2.north) node[circ]{};
      \draw [line, rounded corners=2, thick] (Disc2) |- (AND.in 1);
      \draw [line, rounded corners=2, thick] (Disc3) |- (AND.in 2);
      \draw [line, rounded corners=2, thick] (AND.out) -- (Or2.in 2);
      \draw [line, rounded corners=2, thick, color=red, thick] (Or2.out) -- (FADC.west |- Or2.out) node[circ]{};
      \draw [line, rounded corners=2, thick] (NaI.east) -- +(1.8, 0) |- (Disc3);
      \draw [line, rounded corners=2, thick] (NaI.east) -- +(1.8, 0) |- ($(FADC.west)+(0, -3)$) node[circ]{};

      \node at ($(FADC.west)+(-0.75, -2.4)$) {\texttt{Channel 0}};
      \node at ($(FADC.west)+(-0.75, -3.2)$) {\texttt{Channel 1}};
      \node at ($(FADC.west)+(-0.6, -0.25)$) [color=red] {\texttt{Trigger}};
      \node at ($(Or2.north)+(0, 1.3)$) [color=blue] {\texttt{INHIBIT veto.}};

    \end{circuitikz}
  \end{adjustbox}
  \caption{The DAQ system. The abbreviations R.T., Inh., Disc., and S.A. correspondingly represent random
    trigger, inhibit, discriminator, and shaping amplifier. All of the corsses without node mean non-contact.
    The red input of the Fast Analog-to-Digital Converter (FADC)
    symbolizes the trigger, while the black inputs represent the recording of pulse shapes.
    The inhibit signal (blue line) vetos the current trigger.}
  \label{fig:DAQ}
\end{figure}

\section{Data Analysis}
\label{sec:data-analysis}
\subsection{Energy Calibration}
\begin{figure}[h!tb]
  \subfigure[HPGe energy calibration] {
    \includegraphics[width=0.5\textwidth]{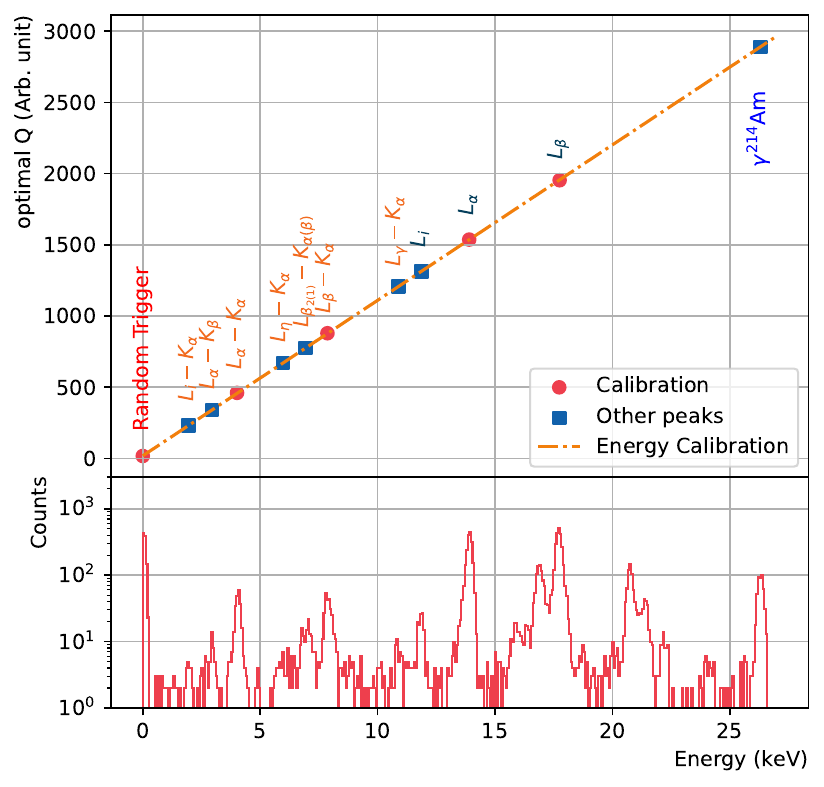}
    \label{subfig:HPGe-calibration}
  }
  \subfigure[NaI{[Tl]} energy calibration] {
    \includegraphics[width=0.5\textwidth]{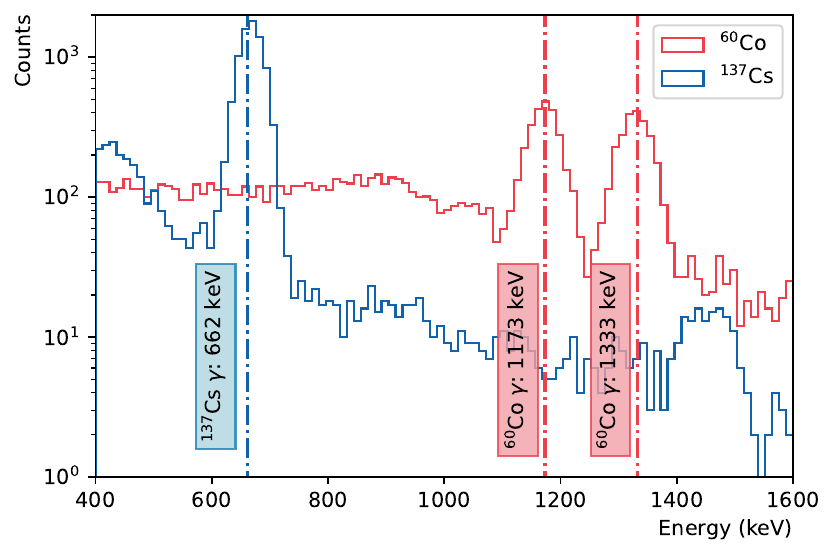}
    \label{subfig:NaI-calibration}
  }

  \caption{Energy calibration for HPGe and  NaI[Tl].
    \ref{subfig:HPGe-calibration} (top)
    displays the linear calibration curve obtained for the
    $^{214}\mathrm{Am}$ source, with blue circles indicating energies used
    in fitting the calibration function, and red squares representing
    reference energy points not used in the fitting.
    The error bars are less than the markers size.
    Data points are labeled according to their source, with random trigger
    events labeled as rad text, $\gamma$ ray from $^{241}\mathrm{Am}$ sub-shells
    then release a Ge characteristic X-ray labeled with orange text,
    and $\gamma$ ray from $^{241}\mathrm{Am}$ sub-shells labeled with blue text.
    The energies diviations are within 50 eV.
    (bottom) The calibration energy spectrum for $^{214}\mathrm{Am}$
    source.
    \ref{subfig:NaI-calibration} illustrates three
    peaks form $^{137}\mathrm{Cs}$ and $^{60}\mathrm{Co}$ used in linear energy clibration.
    Zero-energy is obtained from the random triggers.}
  \label{fig:calibration}
\end{figure}
Two detectors were calibrated individually.
As illustrated in Fig.~\ref{subfig:HPGe-calibration}, the HPGe detector was
calibrated using pulse integral from -4 to 12 $\mathrm{\mu s}$ for
$\mathrm{SA}_2$ and from -12 to 36 $\mathrm{\mu s}$ for $\mathrm{SA}_6$ with
a $^{241}\mathrm{Am}$ X-ray source.
The deviation and non-linearity of calibrated energy do not exceed
50 eV and 0.2\%, respectively.
The NaI[Tl] detector is calibrated using $^{60}\mathrm{Co}$ and
$^{137}\mathrm{Cs}$, as illustrated in Fig.~\ref{subfig:NaI-calibration}.
For both the HPGe and NaI[Tl] detectors, the zero-energy point is obtained
via random trigger events.
At the beginning of each measurement, we performed energy calibration for
both detectors.

\begin{figure}[]
  \centering
  \subfigure[Pedestal selection] {
      \includegraphics[width=0.5\textwidth]{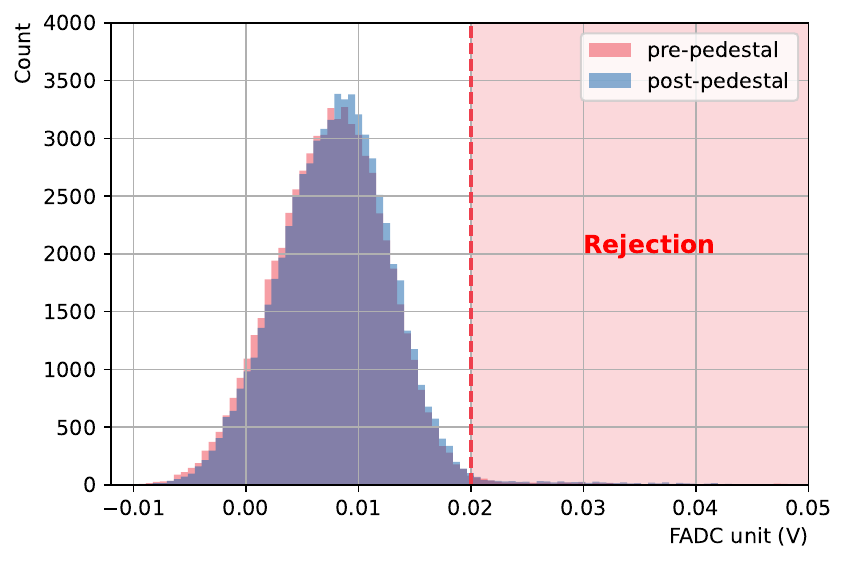}
      \label{subfig:ped-cut}
  }
  \subfigure[A-E and PSD selection]{
      \includegraphics[width=0.5\textwidth]{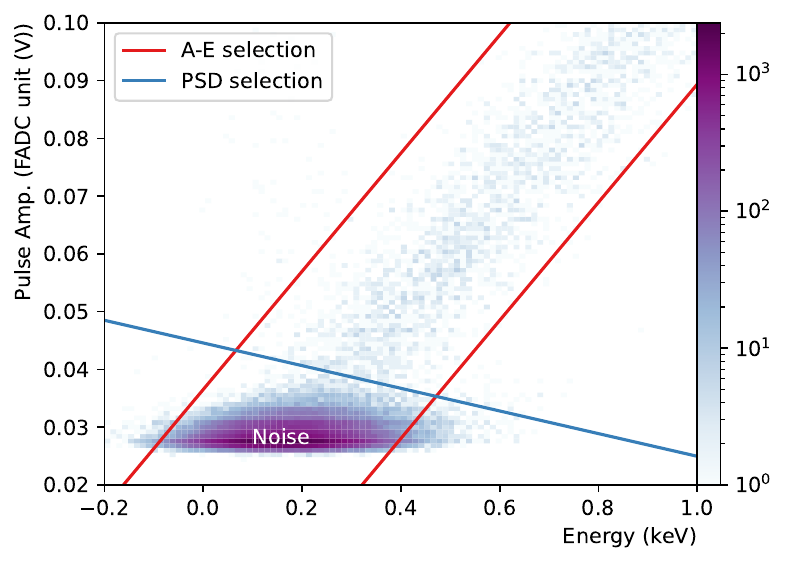}
      \label{subfig:qa-psd-cut}
  }
  \subfigure[Compton candidate selection]{
      \includegraphics[width=0.5\textwidth]{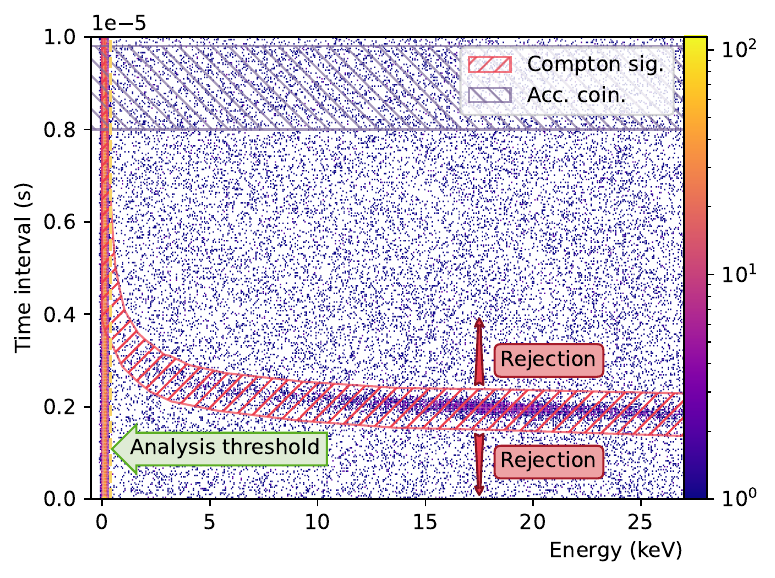}
      \label{subfig:candidate-selection}
  }

  \caption{The illustration of Compton candidate events selection in
    different parameter space.
    \ref{subfig:ped-cut}, \ref{subfig:qa-psd-cut} and
    \ref{subfig:candidate-selection} represent for pedestal selection,
    A-E and PSD selection as well as Compton coincidence events
    selection, respectively. Details are disscussed in Section.~\ref{subsec:compton-candidate-selection}.}
  \label{fig:data-flow}
\end{figure}

\subsection{Compton Candidates Selection}
\label{subsec:compton-candidate-selection}
Candidate events were selected using the following criteria.

\textbf{HPGe reset removal.}
A timing-definite noise structure is introduced due to resetting the
HPGe pre-amplifier.
The HPGe INHIBIT output tags this noise.
We removed the period of 20 $\mathrm{\mu s}$ after the INHIBIT signal was
triggered.

\textbf{Pedestal selection.}
The improper pedestal events lead to inaccuracy energy.
The pre-pedestal and the post-pedestal are definded as
mean amplitudes in the leading and ending 8 $\mu s$ of a pulse shape.
Then, these events are removed via a pedestal cut (see Fig.~\ref{subfig:ped-cut}).

\textbf{A-E selection.}
Multiple trigger events, i.e., events where more than one event triggers
off the HPGe within a time window, can lead to inaccuracy energy.
These multiple trigger events can be effectively removed in the
Amplitude-Energy (A-E) parameter space, as illustrated in
Fig.~\ref{subfig:qa-psd-cut}.
We assume E obeys Gaussian distribution and reject events
beyond the $3\sigma$ region.

\textbf{Coincidence events selection.}
The Compton events trigger the HPGe and NaI[Tl] simultaneously.
Due to the shaping time, the coincident events should occur within a 2
$\mathrm{\mu s}$ (or 6 $\mathrm{\mu s}$) trigger time interval.
We utilize a selection band to identify coincident events, as illustrated
in Fig.~\ref{subfig:candidate-selection}.
The selection band rises in the low-energy region, since the low-energy
events take longer to access the energy threshold in the leading edge
discriminator.

The selection efficiencies are summarized in
Table~\ref{tab:basic-cut-efficiency}.
The corresponding corrections have been applied to the final energy
spectrum for all measurements.
Among these, the DAQ dead time, reset time removal, and improper
pedestal efficiencies are estimated by comparing with the number of random
triggers.
The efficiency of the A-E selection is set at 99.7\% based on the
$3\sigma$ criterion.
The Compton signal band selection efficiency was estimated using Gaussian
fitting, and a slightly conservative estimation is adopted in this work.
Besides, there is an energy-dependent efficiency correction in the
near-threshold region, which will be discussed individually in
Section~\ref{sec:low-ene-eff}.

\begin{table}[t]
  \caption{The energy independent Compton candidates selection efficiency.}
  \label{tab:basic-cut-efficiency}
  \renewcommand{\arraystretch}{1.3} 
  \begin{ruledtabular}
    \begin{tabular}{lcccccc}
\multirow{2}{*}{Selections} & \multicolumn{6}{c}{Efficiency (\%)}           \\ \cline{2-7}
                            & $1.5^\circ$   & $2^\circ$     & $3^\circ$     & $4^\circ$     & $5^\circ$     & $12^\circ$    \\ \hline
DAQ dead time               & 99.9 & 99.9 & 99.9 & 98.2 & 98.1 & 98.0 \\
Reset time                  & 99.6 & 99.8 & 99.8 & 99.8 & 99.8 & 99.8 \\
Pedestal                    & 95.6 & 99.6 & 99.5 & 99.6 & 99.3 & 99.6 \\
A-E selection               & 99.7  & 99.7  & 99.7  & 99.7  & 99.7  & 99.7  \\
Events Selection            & 98.7  & 99.3  & 99.3  & 99.3  & 99.3  & 99.3  \\ \hline
Total                       & 93.6 & 98.3  & 98.2  & 96.6  & 96.3  & 96.4
\end{tabular}
  \end{ruledtabular}
\end{table}

\subsection{Efficiency Correction in the Low-energy Region}
\label{sec:low-ene-eff}

One experimental goal is to measure CS at a small
scattering angle to discriminate between scattering functions and models.
However, the electronic noise of the HPGe detector contaminates the
low-energy region where the Compton peak is located for small scattering
angle measurements.
To mitigate noise leakage, a pulse shape discrimination (PSD) selection is
applied in the A-E parameter space,
as illustrated in Fig.~\ref{subfig:qa-psd-cut}.

\begin{figure}[tpb]
  \centering
  \includegraphics[width=0.5\textwidth]{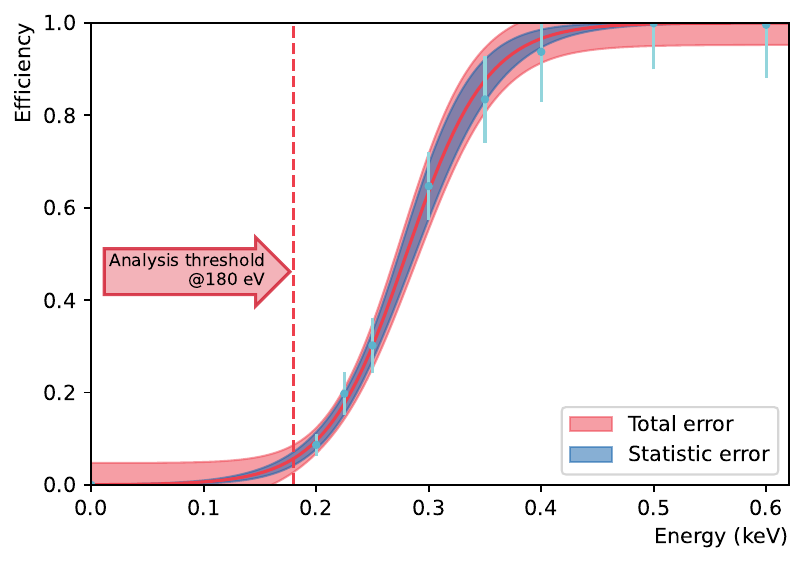}

  \caption{Low-energy efficiency of PSD cut with $1\sigma$ error regions.
    Blue region is statistic error raised by $\arctan$ function fit, while
    red one is the total error including the systematic error arising from
    up-down shift.
    Energy of half-efficiency is $0.28 \mathrm{keV}$.}
  \label{fig:psd-eff}
\end{figure}

The low-energy PSD efficiencies estimation are performed by fitting
PSD-selected events in narrow energy regions with a Gaussian
distribution.
The PSD efficiency for a given energy is defined as the ratio of residual
events to the total events predicted by the Gaussian.
The efficiencies vary as a function of HPGe energy, which can be
well-described by the arctangent function.
The fitting results, and $1\sigma$ uncertainty band are
illustrated in Fig.~\ref{fig:psd-eff}.

\subsection{Background Removal}
\label{sec:background-removal}

\begin{figure}[t]
  \centering
  \includegraphics[width=\linewidth]{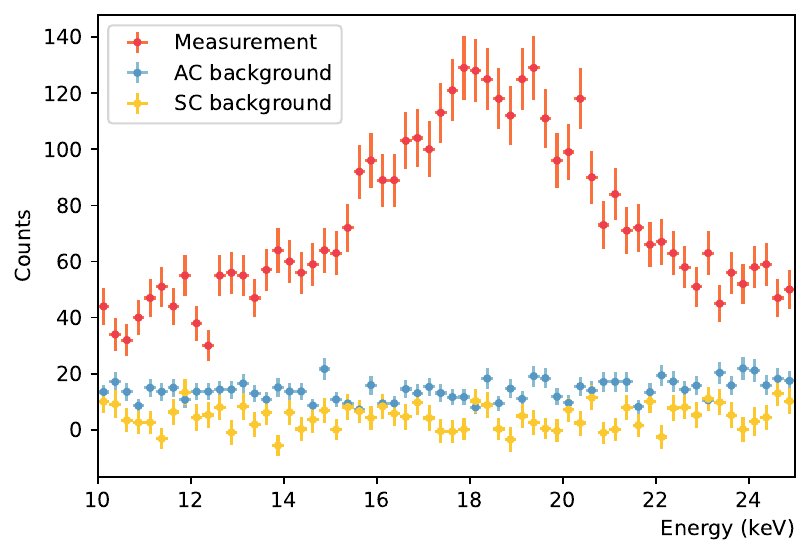}
  \caption{Background removal illustration at the scattering angle of
    $12^\circ$.
    The red crosses are the raw data selected through coincidence
    selection (AC is abbreviate form accidental coincidence).
    The blue crosses are estimated accidental coincidence events.
    And yellow corsses are estimated source-correlated background events
    (SC is abbreviate form source-correlated).
  }
  \label{fig:bkg-removal-demo}
\end{figure}
The preliminary selected events via signal band in
Section~\ref{subsec:compton-candidate-selection}
include Compton signals and backgrounds.
As Fig.~\ref{fig:bkg-removal-demo} illustrated,
we classify backgrounds into two categories: accidental coincidence and
source-correlated.

One is accidental coincidence events by triggering HPGe and NaI[Tl]
detectors within time widows due to environmental radioactivity.
The distribution of time intervals of accidental coincidence events is
nearly uniform (they should follow an exponential distribution, in our
situation, it can be approximated as a uniform distribution).
Thus, we can set an accidental coincidence band mutually exclusive
with the signal band (Fig.~\ref{subfig:candidate-selection})
and then normalize it onto the signal band to
estimate the accidental coincidence background level.

The amount of source-correlated background accounts for approximately 20\%
of the accidental coincidence background, exhibiting energy dependence.
This source-correlated background is identified as contributions from
$\gamma$ ray with energies lower than the 662 keV from $^{137}\mathrm{Cs}$
due to the shielding effect of the lead collimator.
This background was observed when selecting energy ranges for the NaI[Tl]
detector between 150 and 750 keV, while it was dismissed in the ranges of
600 to 700 keV and above 700 keV.
This suggests that the source of this background arises from gammas below
662 keV.
To validate this, we simulated the collimator influence for
$^{137}\mathrm{Cs}$ source.
The simulation results indicate that the Pb-collimator will lead to
approximately 14.8\% of the incident gammas having energies lower than the
characteristic peak of $^{137}\mathrm{Cs}$.
The trend of the energy dependence of the simulated backgrounds is
consistent with the experimental observations.

However, due to the computational expensiveness, the correlated
backgrounds are not estimated through simulations.
We removed these backgrounds by subtracting normalized correlated
backgrounds from non-signal energy regions at other scattering angles
accordingly.
This energy-dependent correlated-backgroud would change the shape of the
energy spectrum.
Through this removal, our previous data show better agreement with the
simulations than the earlier
results~\cite{jiaHighaccuracyMeasurementCompton2022}.

%


%
\subsection{Systematic Errors}
\label{sec:systematic-errors}
\begin{table*}[t]
  \caption{Systematic errors in measurements. Two kinds of systematic
    errors are listed by categories.
    The serial numbers represents for items discussed in
    Section.~\ref{sec:systematic-errors}.
    Errors from the low-energy efficiency and statistical error in
    simulations are estimated together for $1.5^\circ$ and $2^\circ$.
  }
  \label{tab:systematic-error}
  \begin{ruledtabular}
    \renewcommand{\arraystretch}{1.3} 
    \begin{tabular}{ccccccccccc}
\multicolumn{1}{l}{\multirow{2}{*}{Scattering angles}} &
  \multicolumn{5}{c}{\begin{tabular}[c]{@{}c@{}}Energy Dependent \\ Systematic Errors (\%)\end{tabular}} &
  \multicolumn{4}{c}{\begin{tabular}[c]{@{}c@{}}Energy Independent \\ Systematic Errors (\%)\end{tabular}} &
  \multicolumn{1}{l}{\multirow{2}{*}{\begin{tabular}[c]{@{}l@{}}Total Systematic Error\\ (\%)\end{tabular}}} \\ \cline{2-10}
\multicolumn{1}{l}{} &
  \multicolumn{1}{l}{(a)} &
  \multicolumn{1}{l}{(b)} &
  \multicolumn{1}{l}{(c)} &
  \multicolumn{1}{l}{(d)} &
  \multicolumn{1}{l}{total} &
  \multicolumn{1}{l}{(a)} &
  \multicolumn{1}{l}{(b)} &
  \multicolumn{1}{l}{(c)} &
  \multicolumn{1}{l}{total} &
  \multicolumn{1}{l}{} \\ \hline
$1.5^\circ$ & 0.23 & 0.00 & \multicolumn{2}{c}{0.67} & 0.71 & 0.01 & 0.03 & 0.06 & 0.07 & 0.72 \\
$2^\circ$   & 0.22 & 0.00 & \multicolumn{2}{c}{0.70} & 0.77 & 0.01 & 0.04 & 0.06 & 0.07 & 0.77 \\
$3^\circ$   & 0.37 & 0.00 & -         & 0.31         & 0.49 & 0.01 & 0.06 & 0.07 & 0.09 & 0.50 \\
$4^\circ$   & 0.34 & 0.00 & -         & 0.29         & 0.45 & 0.01 & 0.05 & 0.07 & 0.08 & 0.46 \\
$5^\circ$   & 0.40 & 0.01 & -         & 0.28         & 0.49 & 0.02 & 0.08 & 0.09 & 0.12 & 0.51 \\
$12^\circ$  & 0.43 & 0.01 & -         & 0.29         & 0.51 & 0.01 & 0.04 & 0.07 & 0.09 & 0.52
\end{tabular}
\end{ruledtabular}
\end{table*}
This work considers two categories of systematic errors, namely
energy-dependent and energy-independent.
The energy-dependent part affects the shape of the energy spectra, while
the energy-independent part only affects the total amount of events.
The systematic errors are listed according to their energy dependence, as
presented in the following items.
And all of the systematic errors are summarized in
Table~\ref{tab:systematic-error}.

The energy-dependent systematic errors can be categorized as follows:

\textbf{(a) Error of the scattering angle}.
This error arises from three processes: calibration of the horizontal
plane, zero-angle determination, and arbitrary scattering angle
calibration, as discussed in~\ref{subsec:angle-clibration}.
The error in the calibration of the horizontal plane is influenced by the
laser level.
It is estimated by placing laser level on the calibrated plane and
rotating it $90^\circ$ several times.
The width of the laser at 5 m (same as horizontal plane calibration) is
4 mm.
This could result in a maximum error of $0.02^\circ$ in the horizontal plane,
which is consistent with a manufacturer-provided value of
$0.03\ \mathrm{mm/m}$.
The error associated with the zero scattering angle distribution is
assigned a value of $0.01^\circ$ based on the two measurements discussed in
Section~\ref{subsec:angle-clibration}.
The scattering angle calibration introduces a systematic error of
$0.02^\circ$.
We attribute this error to the 1 mm half-width of the laser dislocating
the center of the NaI[Tl] detector, resulting in a conservative estimate
(1 mm displacement at 2 m).
The overall contribution of the scattering angle error is less than
$0.03^\circ$.

\textbf{(b) Indicator deformation}.
The deformation of the stainless-steel indicator in the lead collimator
leads to an increased influx of $\gamma$ rays with energies below 662 keV
into the detector, consequently affecting the energy spectra.
The laser level limited the indicator deformation to a value less than
$0.3^\circ$.
By applying a conservative $0.3^\circ$ offset in the simulation,
the results indicate a mere $0.2\%$ discrepancy in the spectra.

\textbf{(c) Low-energy efficiency correction}.
As illustrated in Fig.~\ref{fig:psd-eff}, the error band of the
efficiency curve is considered as a systematic error in the low-energy
efficiency correction.
This uncertainty comprises two parts: the first is the estimation
uncertainty associated with the least squares method used to fit the
two-parameter hyperbolic tangent function, and the second is the
systematic error arising from the choice of function.
In our case, we introduced an additional parameter to describe the
up-and-down shift, which addresses the error associated with the choice
of functions.

\textbf{(d) Statistical error of simulations}.
The statistical error of simulations is regarded as a systematic error
when comparing the simulated spectra with mearsured energy spectra.
This error is typically considered minor, as simulations usually
generate a sufficient number of events to minimize statistical
fluctuations.

In addition, the energy-independent systematic errors can be categorized
as follows:

\textbf{(a) Normalization factor}.
A CS model has to be chosen to normalize the simulations
to the experiments.
Although different models exhibit varying DDCS behavior in the low-energy
region, they share the same DCS\@.
This implies that a broad energy region results in a reduced impact from
the choice of models.
The uncertainty of the normalization factor is derived from two models
that differ in the number of events within a specific energy region,
particularly from 1 to 60 keV at a scattering angle of 12 degrees.
This item introduces a 0.09\% error on the normalization factor.

\textbf{(b) Efficiency estimations}.
The error associated with the energy-independent efficiency correction is
considered one of the systematic errors.
This error arises from the statistical fluctuations of random trigger
events.

\textbf{(c) Multiple trigger}.
During the experiment, multiple particles may incidentally strike the
HPGe detector simultaneously, leading to their erroneous identification as
a single experimental signal.
Through pulse shape analysis, it was determined that the number of events
triggered by two signals within an 8 $\mathrm{\mu s}$interval was only
0.5\%.

All of the systematic errors have been incorporated into the simulated
energy spectra.

\begin{table*}[tbp]
    \caption{The experimental scattering angles, effective scattering angles, corresponding experimental X.
    Besides, $\chi^2$ statistic test for Livermore, Monash models and
    scattering angle of $1.5^\circ$, $2^\circ$, $3^\circ$, $4^\circ$, $5^\circ$ and $12^\circ$.}
  \label{tab:models-hypo-test}
  \begin{ruledtabular}
    \renewcommand{\arraystretch}{1.3} 
    \begin{tabular}{lllllllll}
     \multicolumn{1}{c}{\multirow{2}{*}{Exp.$^{a}$}} &
  \multirow{2}{*}{Eff.$^{b}$} &
  \multirow{2}{*}{$\bar X^{c}$} &
  \multicolumn{2}{c}{Chisquare} &
  \multicolumn{1}{c}{\multirow{2}{*}{ndf.$^{d}$}} &
  \multicolumn{2}{c}{$p$-value} &
  \multirow{2}{*}{Z$^{e}$} \\ \cline{4-5} \cline{7-8}
\multicolumn{1}{c}{} &       &      & Livermore & Monash & \multicolumn{1}{c}{} & Livermore & Monash               &      \\ \hline
12                   & 11.97 & 5.56 & 51.84     & 163.69 & 59                   & 0.734     & $8.79\times10^{-12}$ & 6.72 \\
5                    & 5.03  & 2.34 & 55.50     & 139.15 & 59                   & 0.605     & $2.00\times10^{-8}$  & 5.49 \\
4                    & 4.06  & 1.89 & 73.64     & 184.74 & 79                   & 0.649     & $1.88\times10^{-10}$ & 6.26 \\
3                    & 3.23  & 1.50 & 83.10     & 192.46 & 79                   & 0.354     & $1.86\times10^{-11}$ & 6.61 \\
2                    & 2.34  & 1.08 & 104.42    & 226.42 & 85                   & 0.075     & $8.51\times10^{-15}$ & 7.67 \\
      1.5                  & 2.02  & 0.94 & 113.67    & 170.21 & 97                   & 0.119     & $6.34\times10^{-6}$  & 4.37 \\ \hline
      Total                & --     & --    & 482.17    & 1076.67& 458                  & 0.210     & $8.996\times10^{-52}$& --

\end{tabular}
\end{ruledtabular}

    \leftline{$^{a}$Experimental scattering angle.}
    \leftline{$^{b}$Effective scattering angle.}
    \leftline{$^{c}$Experimental effective $X$.}
    \leftline{$^{d}$Number of degree of freedom.}
    \leftline{$^{e}$Equivalent significance of a discrepancy between the
      data and Monash model.}
\end{table*}

\section{Results and Discussion}
\label{sec:results-discussion}

\subsection{Doubly Differential Cross-Section}
\label{sec:ddcs}

\begin{figure*}
  \centering
  \subfigure[] {
    \includegraphics[width=0.48\textwidth]{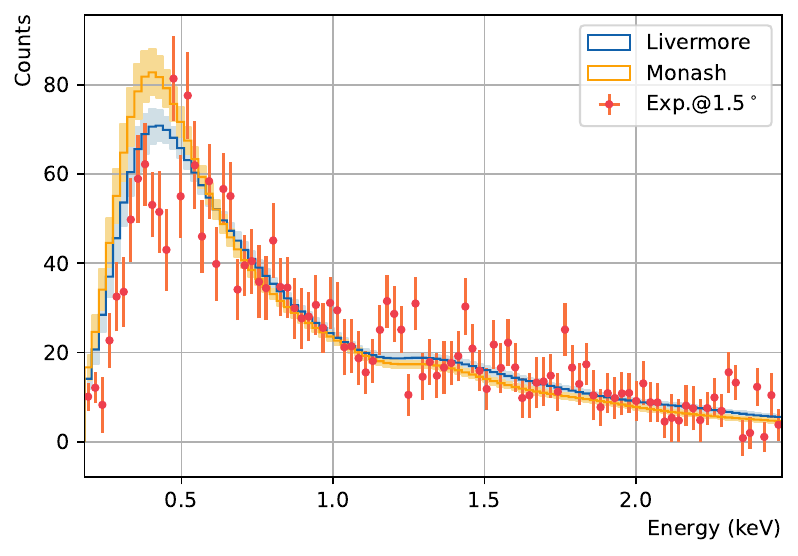}
    \label{fig:spect-1.5deg}
  }
  \subfigure[] {
    \includegraphics[width=0.48\textwidth]{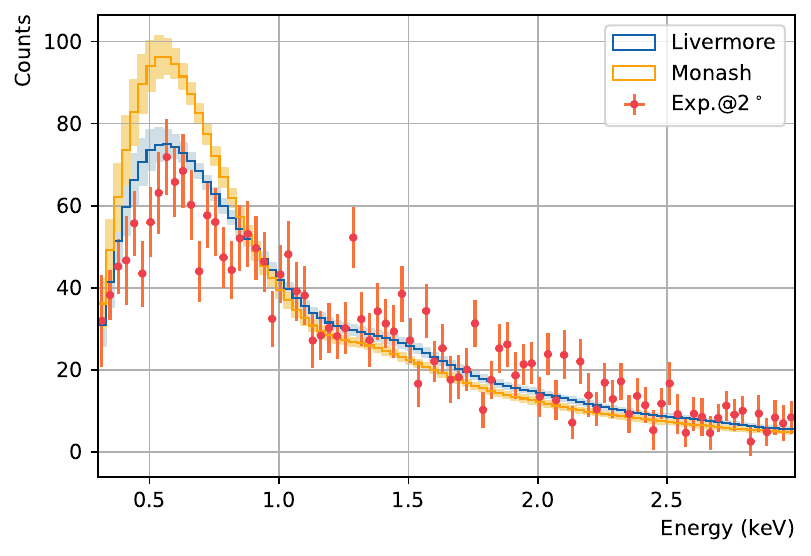}
    \label{fig:spect-2deg}
  }
  \subfigure[] {
    \includegraphics[width=0.48\textwidth]{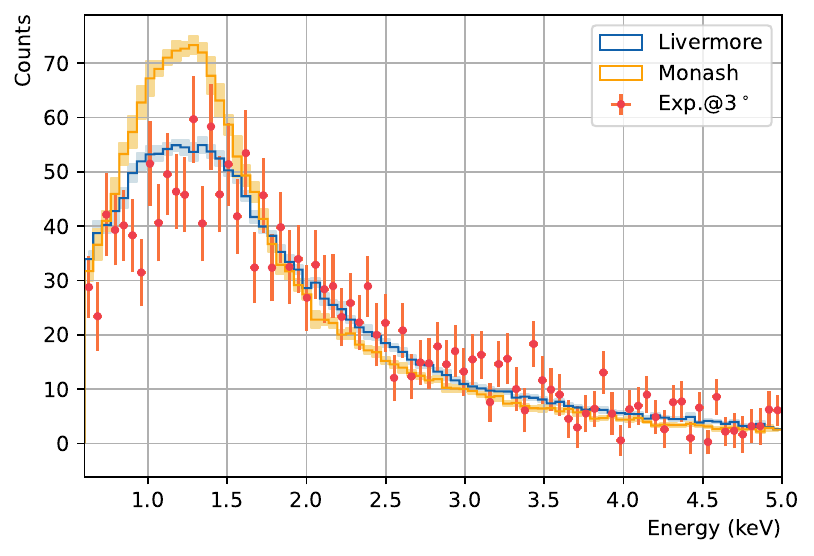}
    \label{fig:spect-3deg}
  }
  \subfigure[] {
      \includegraphics[width=0.48\textwidth]{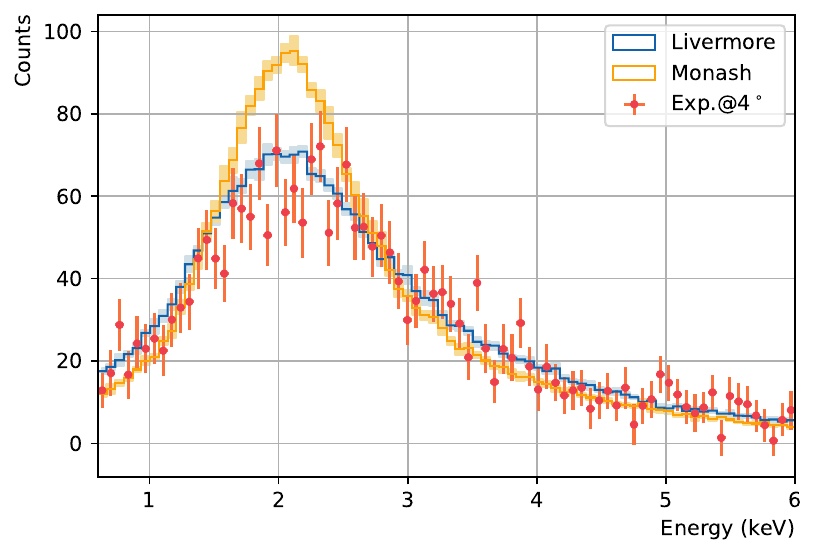}
      \label{fig:spect-4deg}
  }
  \subfigure[] {
      \includegraphics[width=0.48\textwidth]{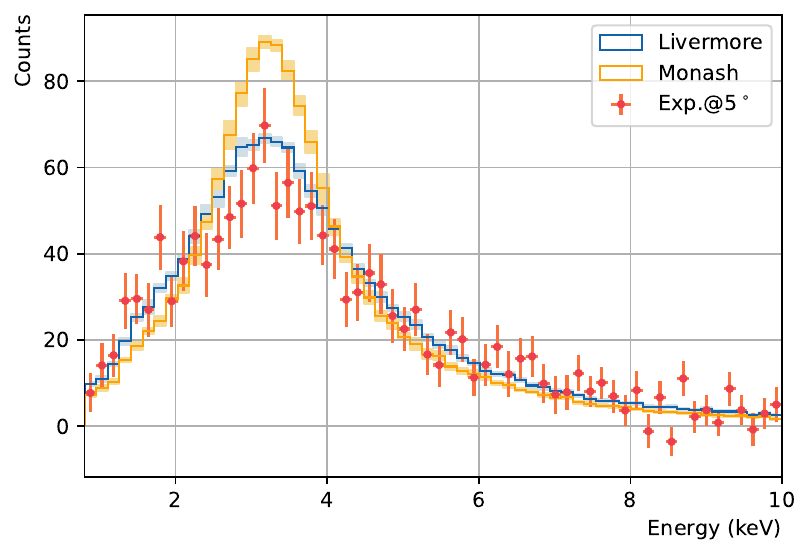}
      \label{fig:spect-5deg}
  }
  \subfigure[] {
      \includegraphics[width=0.48\textwidth]{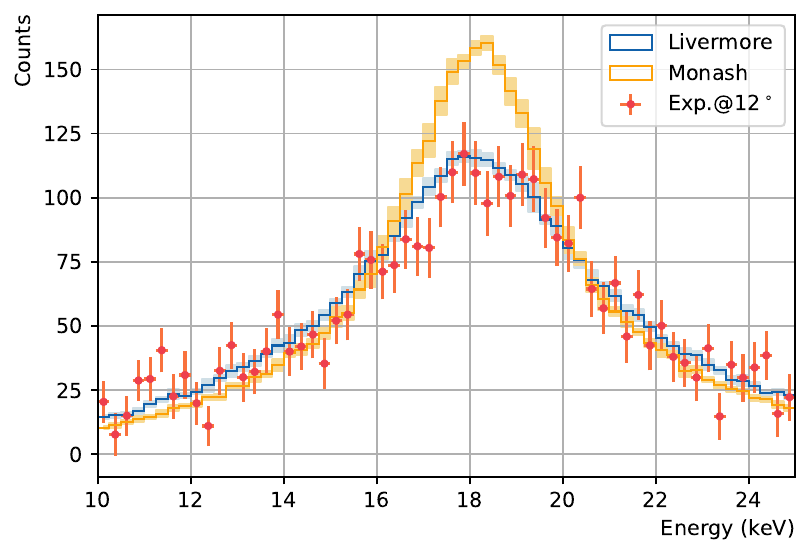}
      \label{fig:spect-12deg}
  }
  \caption{Measured energy spectra (rad cross) for several scattering
    angles vary form $1.5^\circ$ to $12^\circ$ as well as its comparison between
    \geant\@models.
    Figures~\ref{fig:spect-1.5deg} to \ref{fig:spect-12deg} correspond to
    the scattering angle of $1.5^\circ, 2^\circ, 3^\circ, 4^\circ, 5^\circ$ and $12^\circ$,
    respectively.
    The Livermore (in blue) and Monash (in orange) models are depicted
    as histograms after applying efficiency corrections.
    The systematic errors for each angle are shown as error bands
    (color shadows), respectively.
  }
  \label{fig:ene-spct}
\end{figure*}

The energy spectra corresponding to scattering angles of $1.5^\circ$, $2^\circ$,
$3^\circ$, $4^\circ$, $5^\circ$ and $12^\circ$ were measured.
To compare the measurememts with our calculations, we reconstructed the
experimental geometries in \geant\@simulations to account for the
influence of geometry, source dispersion and detector response on the
spectra.

The simulated spectra are normalized to the measured spectra.
The normalization factor was determined through the count rate from 1 to
60 keV at $12^\circ$, aiming to significantly suppress the discrepancies in
DDCS introduced by the \geant\@Compton models.
Based on this, we established a correlation between the number of events
generated by simulation and the DAQ period.
We then applied this normalization factor to the other measurements at
different scattering angle, in accordance with their respective
measurement times.

Measured and corresponding simulated energy spectra for scattering angles
from $1.5^\circ$ to $12^\circ$ are illustrated in Fig.~\ref{fig:ene-spct}.
The measured spectra (red crosses) are presented after background
removal, and the simulated spectra (step histogram) are presented after
efficiency correction as well as systematic error association
(shadow region).
Our measurements are capable of distinguishing between these two models.
We perform the Pearson’s chi-square hypothesis test between simulated and
measured spectra.
The corresponding results are summarized in
Table~\ref{tab:models-hypo-test}.
Due to refined data processing and background removal, our results
significantly reduce the inconsistency with the Livermore model when
compared to previous measurements reported in
Ref.~\cite{jiaHighaccuracyMeasurementCompton2022}.

The accuracy of the Livermore (Penelope) model is validated across most
scattering angles.
In contrast, the data strongly refutes the Monash model for all scattering
angles.
The equivalent significance of the discrepancy between the data and Monash
model
ranges from $5.49\sigma$ to $7.67\sigma$ for scattering angles between $2^\circ$ and
$12^\circ$.
Although, the significance remains substantial for the $1.5^\circ$ data,
it does not surpass the $5\sigma$ criterion.
This limitation can be attributed to the efficiency correction in the
low-energy region, which smears out the most critical area,
namely the Compton peak region.

However, as momentum transfer decreases, even the Livermore models
gradually lose consistency with the experimental spectra, as
illustrated in Figures~\ref{fig:spect-1.5deg} and \ref{fig:spect-2deg}.
The peak region for small-angle measurements falls into the sub-keV range,
indicating that only outer-shell electrons ($\mathrm{i.e.}$, covalent
electrons) are excited as final states.
The structure of covalent electrons in Ge differs from that of isolated
atomic systems due to solid-state effects.
The observed weak consistency suggests that viewing outer-shell electrons
from the perspective of isolated atoms in the Compton model within this
region is inaccurate.
Instead, the influence of the solid-state system (covalent crystal) on
electronic structure and momentum distribution should be included.

In this work, we carried out two new measurements at the scattering angles
of $1.5^\circ$ and $5^\circ$, respectively.
Additionally, we refined the remaining data using more sophisticated data processing methods.
The measurements favor the Livermore model and decisively reject the
Monash model, which is consistent with the qualitative findings of
previous studies~\cite{jiaHighaccuracyMeasurementCompton2022}.
In the following work, we adopt the Livermore model in the SF
normalization as well as dark matter background simulations.

\subsection{The Scattering Function}
\label{sec:scattering-function}

\begin{figure}[tb]
  \centering
  \includegraphics[width=\linewidth]{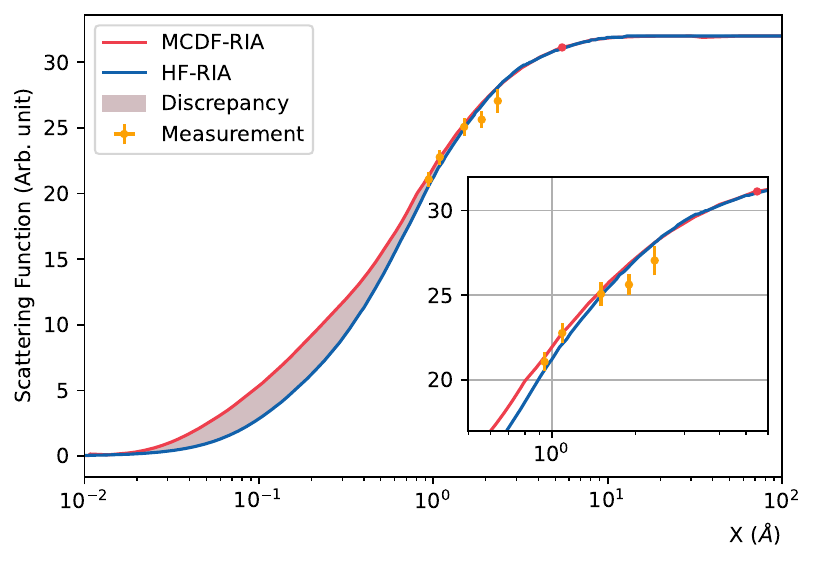}
  \caption{The scattering functions obtained from HF-RIA by
    \citet{hubbellAtomicFormFactors1975} (blue line) and our MCDF-RIA
    results (red line), compared with experimental measurements at the
    scattering angle (effective scattering angle) of $1.5^\circ (2.02^\circ)$,
    $2^\circ(2.34^\circ)$, $3^\circ(3.23^\circ)$, $4^\circ(4.06^\circ)$, $5^\circ(5.03^\circ)$ and $12^\circ$.
    Measurement at $12^\circ$ (red dot) is regraded as calibration point.
    The error of $X$ is assigned but smaller than the marker size.}
  \label{fig:scattering-fucntion}
\end{figure}

The defination of the measured scattering function is given by
\begin{equation}
  \label{eq:exp-sf}
  {S(\bar{X})}_\mathrm{exp.} = \left[\left(\frac{d\sigma}{d\Omega}\right)_\mathrm{exp.} \Big/ \left(\frac{d\sigma}{d\Omega}\right)_\mathrm{sim.}\right]\cdot S(\bar X)_\mathrm{sim.},
\end{equation}
where the subscripts ``$\mathrm{exp.}$'' and ``$\mathrm{sim.}$'' denote
experimental measurements and the theoretical SF using in simulation,
respectively.
The symbol $\bar X$ represents for the effective $X$ of a measurement,
which will be discussed further below.
As mentioned in Section~\ref{sec:ddcs}, the normalization from the
simulated spectra to measurements is established through calibration at
$12^\circ$, indicating that the SF at $12^\circ$ aligns precisely with the
theoretical SFs (corresponding to the red dot in
Fig.~\ref{fig:scattering-fucntion}).

Due to source dispersion, geometric effects, and low-energy efficiencies,
the actual scattering angle does not have a fixed value but follows a
distribution centered around the experimental setup.
We obtained the scattering angle distributions from simulations and
calculated the corresponding $\bar X$ through the mean value of its SF
distributions.
Consequently, the effective scattering angles are larger than those in
experimental configuration, and the severity of this pathology increases
as the experimental scattering angle decreases.
When investigating the scattering behavior towards an even lower momentum
transfer region, a ``soft wall'' emerges.
The effective scattering angle is $2.02^\circ$ for an experimental scattering
angle of $1.5^\circ$, while our successive simulations at a scattering angle
of $1^\circ$ reveal an effective scattering angle of $2.17^\circ$.
This finding has limited our ability to investigate lower scattering
angles, leading us to fix our measurements at the smallest angle at
$1.5^\circ$.
The relation of experimental and effective scattering angles is listed
in Table~\ref{tab:models-hypo-test}.

In Fig.~\ref{fig:scattering-fucntion}, we present measured SFs (orange
crosses) alongside our MCDF-RIA calculation (red line) and the results of
\citet{hubbellAtomicFormFactors1975} (blue line).
The measured data are consistent with both SFs within the error bars in
high momentum transfer regions where the theoretical predictions align.
However, in the low momentum transfer region, the significance of the
discrepancy increases as momentum transfer decreases.
In this context, the experimental results cannot exclude the scattering
functions but favor the MCDF-RIA scattering function.

As discussed perviously, the emergence of the ``soft wall'' phenomenon
makes it challenging to investigate the lower momentum transfer region for
enhanced discrimination capability with the current experimental setup.
Further exploration in this area may necessitate an updated experimental
approach.

\section{Background of Electronic Recoil Channel}
\label{sec:compt-scatt-DM-bkg}

\subsection{Compton Background Evaluation}
\label{sec:backgr-electr-reco}

\begin{table*}[t]
  \caption{The height ratio of sub-shells in the CS
    background spectra relative to K-shell.
    The configurations represent the detector size, source position, and
    scattering functions employed in the simulations.
    Non-flat Compton steps are indicated with upper index.
  }
  \label{tab:compt-DM-bkg-diff}
  \begin{ruledtabular}
    \renewcommand{\arraystretch}{1.3} 
    \begin{tabular}{llccccccc}
$\gamma$ sources &
  Configurations &
  $L_\mathrm{I}$-$K$ &
  $L_\mathrm{IIa}$-$L_\mathrm{I}$ &
  $M_\mathrm{I}$-$L_\mathrm{IIa}$ &
  $M_\mathrm{IIa}$-$M_\mathrm{I}$ &
  $M_\mathrm{IIIa}$-$M_\mathrm{IIa}$ &
  $N_\mathrm{I}$-$M_\mathrm{IIIa}$ &
  below $N_\mathrm{I}$ \\ \hline
\multirow{5}{*}{$^{212}\mathrm{Pb}$ (239 keV)}  & 1kg,near,MCDF & 0.97 & 0.91 & 0.70 & 0.63 & 0.45 & 0.14 & 0.05 \\
                                                & 1kg,far,MCDF  & 0.97 & 0.91 & 0.74 & 0.67 & 0.48 & 0.15 & 0.04 \\
                                                & 1kg,near,HF   & 0.97 & 0.90 & 0.69$^\S$ & 0.57 & 0.37 & 0.09 & 0.02 \\
                                                & 5g,near,MCDF  & 1.00$^\dagger$ & 0.98 & 0.79 & 0.72 & 0.51 & 0.16 & 0.05 \\
                                                & 5g,far,MCDF   & 1.00$^\dagger$ & 0.98 & 0.78 & 0.72 & 0.50 & 0.18 & 0.04 \\ \cline{2-9}
\multirow{5}{*}{$^{214}\mathrm{Pb}$ (352 keV)}  & 1kg,near,MCDF & 0.95 & 0.89 & 0.71 & 0.65 & 0.46 & 0.14 & 0.03 \\
                                                & 1kg,far,MCDF  & 0.96 & 0.92 & 0.72 & 0.66 & 0.47 & 0.15 & 0.06 \\
                                                & 1kg,near,HF   & 0.96 & 0.89 & 0.69$^\S$ & 0.58 & 0.38 & 0.10 & 0.03 \\
                                                & 5g,near,MCDF  & 0.97 & 0.94 & 0.73 & 0.69 & 0.49 & 0.17 & 0.04 \\
                                                & 5g,far,MCDF   & 0.97 & 0.93 & 0.74 & 0.72 & 0.48 & 0.18 & 0.04 \\ \cline{2-9}
\multirow{5}{*}{\begin{tabular}[c]{@{}l@{}}$\mathrm{e}^{-}$-$\mathrm{e}^+$ annihilation \\(511 keV)\end{tabular}} &
  1kg,near,MCDF &
  0.95 &
  0.89 &
  0.70 &
  0.64 &
  0.45 &
  0.14 &
  0.04 \\
                                                & 1kg,far,MCDF  & 0.95 & 0.90 & 0.71 & 0.65 & 0.46 & 0.15 & 0.04 \\
                                                & 1kg,near,HF   & 0.95 & 0.87 & 0.67$^\S$ & 0.58 & 0.37 & 0.09 & 0.03 \\
                                                & 5g,near,MCDF  & 0.95 & 0.91 & 0.72 & 0.69 & 0.49 & 0.22 & 0.06 \\
                                                & 5g,far,MCDF   & 0.96 & 0.92 & 0.72 & 0.68 & 0.48 & 0.19 & 0.06 \\ \cline{2-9}
\multirow{5}{*}{$^{214}\mathrm{Bi}$ (609 keV)}  & 1kg,near,MCDF & 0.94 & 0.88 & 0.69 & 0.63 & 0.44 & 0.15 & 0.05 \\
                                                & 1kg,far,MCDF  & 0.94 & 0.90 & 0.69 & 0.62 & 0.46 & 0.17 & 0.07 \\
                                                & 1kg,near,HF   & 0.95 & 0.87 & 0.67$^\S$ & 0.57 & 0.36 & 0.12 & 0.03 \\
                                                & 5g,near,MCDF  & 0.96 & 0.94 & 0.71 & 0.73 & 0.51 & 0.29 & 0.06 \\
                                                & 5g,far,MCDF   & 0.96 & 0.92 & 0.72 & 0.71 & 0.48 & 0.25 & 0.05 \\ \cline{2-9}
\multirow{5}{*}{$^{40}\mathrm{K}$ (1461 keV)}   & 1kg,near,MCDF & 0.94 & 0.90 & 0.70 & 0.71 & 0.45 & 0.45 & 0.06 \\
                                                & 1kg,far,MCDF  & 0.95 & 0.92 & 0.72 & 0.72 & 0.45 & 0.45 & 0.05 \\
                                                & 1kg,near,HF   & 0.94 & 0.89 & 0.68$^\S$ & 0.66 & 0.40 & 0.40 & 0.06 \\
                                                & 5g,near,MCDF  & 0.97$^\dagger$ & 1.02 & 0.77$^\dagger$ & 0.83 & 0.62 & 0.62 & 0.15 \\
                                                & 5g,far,MCDF   & 0.95$^\dagger$ & 0.99 & 0.75$^\dagger$ & 0.78 & 0.53 & 0.53 & 0.11 \\ \cline{2-9}
\multirow{5}{*}{$^{208}\mathrm{Tl}$ (2614 keV)} & 1kg,near,MCDF & 0.95 & 0.96 & 0.73 & 0.76 & 0.52 & 0.52 & 0.10 \\
                                                & 1kg,far,MCDF  & 0.94 & 0.94 & 0.72 & 0.74 & 0.47 & 0.47 & 0.06 \\
                                                & 1kg,near,HF   & 0.95 & 0.95 & 0.71$^\S$ & 0.70 & 0.49 & 0.49 & 0.09 \\
                                                & 5g,near,MCDF  & 0.98$^\dagger$ & 1.09 & 0.83$^\dagger$ & 0.93 & 0.77 & 0.77 & 0.24 \\
                                                & 5g,far,MCDF   & 0.97$^\dagger$ & 1.05 & 0.80$^\dagger$ & 0.90 & 0.70 & 0.70 & 0.18
    \end{tabular}
  \end{ruledtabular}
  \leftline{$^\dagger$ Negative slope.}
  \leftline{$^\S$ Positive slope.}

\end{table*}

To provide insights for current and next-generation experiments based on
ionization detection, we first evaluated the impacts of SFs on the
low-energy spectrum, as these cannot be determined experimentally.
Subsequently, we apply our MCDF-RIA calculations to \geant simulations and
investigate two experimental conditions: varying gamma source positions
and detector masses.

The radioactive sources are chosen from the U/Th decay chain.
Elements of the U/Th decay chain are commonly found in rock caves and
materials surrounding detectors, making them a dominant source of gamma
radiation.
In this study, we discuss the gamma rays arising from prevalent
environmental radioactivity, including
$^{212}\mathrm{Pb}$ (239 keV), $^{214}\mathrm{Pb}$ (352 keV),
$^{214}\mathrm{Bi}$ (609 keV), $^{40}\mathrm{K}$ (1461 keV),
$^{208}\mathrm{Tl}$ (2614 keV), as well as those from electron-positron
annihilation (511 keV).

\begin{figure}[tb]
  \centering
  \includegraphics[width=\linewidth]{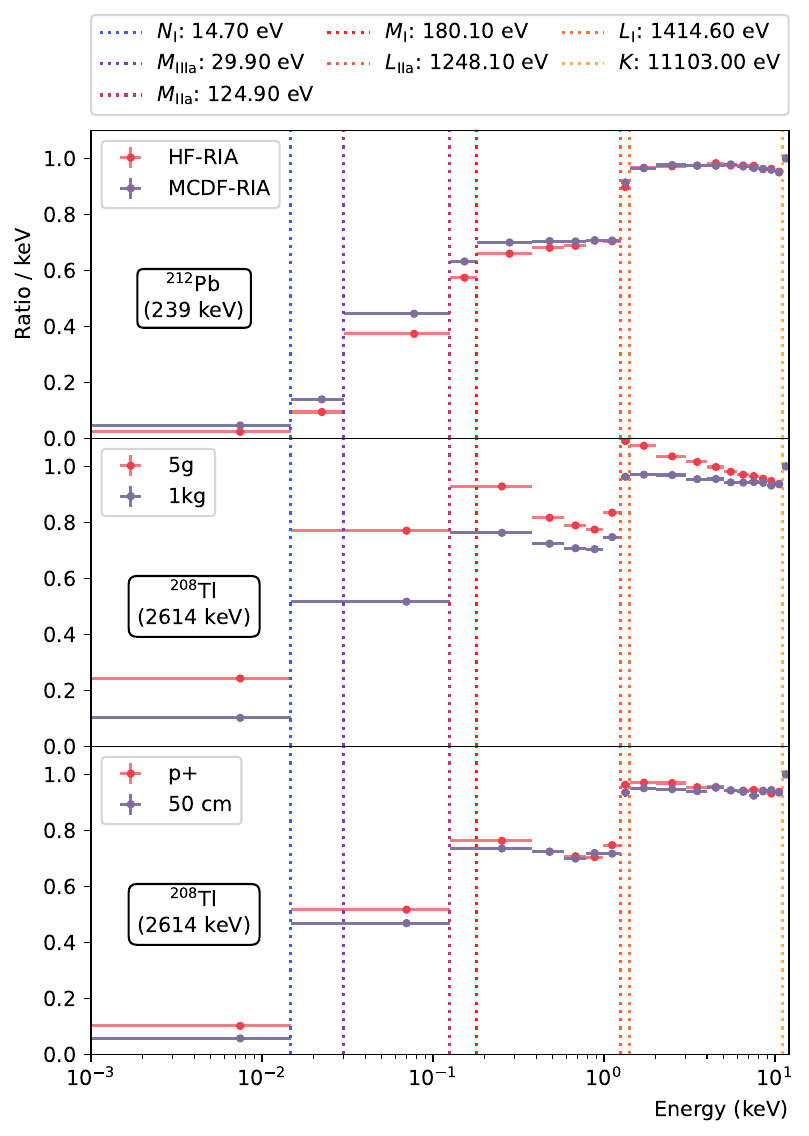}
  \caption{The CS background on the HPGe detector
    for three aspects:
    (top) the scattering functions,
    (middle) the mass of the HPGe detector,
    and (bottom) the position of the gamma sources.
    We demonstrate the spectra with the maximum differences for each
    group.}
  \label{fig:compt-bkg}
\end{figure}

The energy spectra obtained through \geant\@simulations are binned by
sub-shells and normalized to K-shell (the last bin).
Due to the inability of the HPGe experimental spectra to discern event
components, the establishment of the background model heavily relies on
the shape of the background spectrum.
Therefore, we meticulously discussed the shape of the Compton spectra
under various conditions, quantifying the step structure with the height
ratio of the K-shell in the Compton spectrum.
All conditions are compared with a benchmark condition (1 kg HPGe
detector, MCDF-RIA scattering function), where the source is positioned
around the P+ electrode.
The heights of these steps are presented in
Table~\ref{tab:compt-DM-bkg-diff} with non-flat structure highlighted.

We evaluate the difference in the Compton background
energy spectrum under the Livermore model between the recalculated
MCDF-RIA SF and the HF-RIA SF adopted by \textsc{Geant4}.
The steps predicted by the MCDF-RIA SF are approximately 10\% to 50\%
higher than those predicted by the HF-RIA SF at 239 keV.
As illustrated in Fig.~\ref{fig:compt-bkg} (top), the most significance
diffenence appares below $L$-shell ionization energy (sub-keV region).
This discrepancies arise because the MCDF-RIA scattering function is more
pronounced than the previous one in the low-energy transfer region.
This difference gradually diminishes with increasing gamma energy; for
incident gamma rays at 2614 keV, only about a 5\% difference is observed.

\begin{figure}[htp]
  \centering
  \subfigure[] {
    \includegraphics[width=\linewidth]{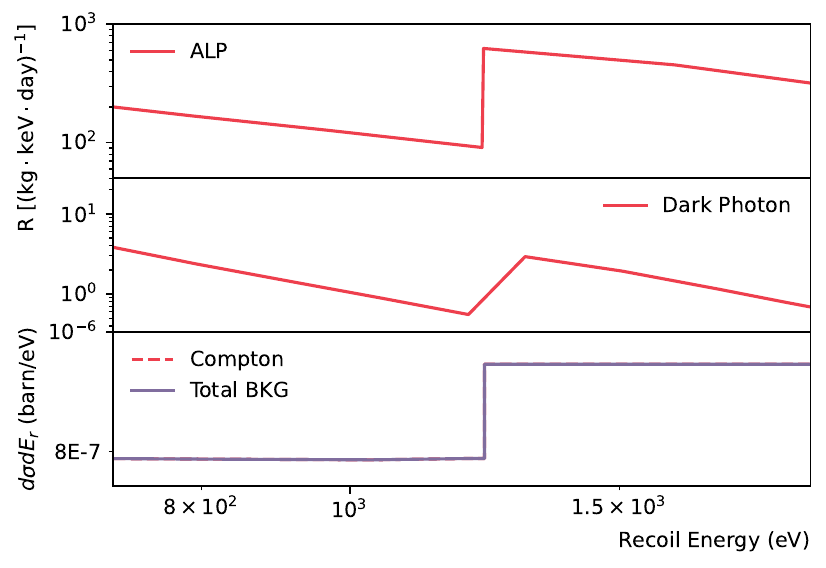}
    \label{fig:alp-dp}
  }
  \vspace{-10pt}
  \subfigure[] {
    \includegraphics[width=\linewidth]{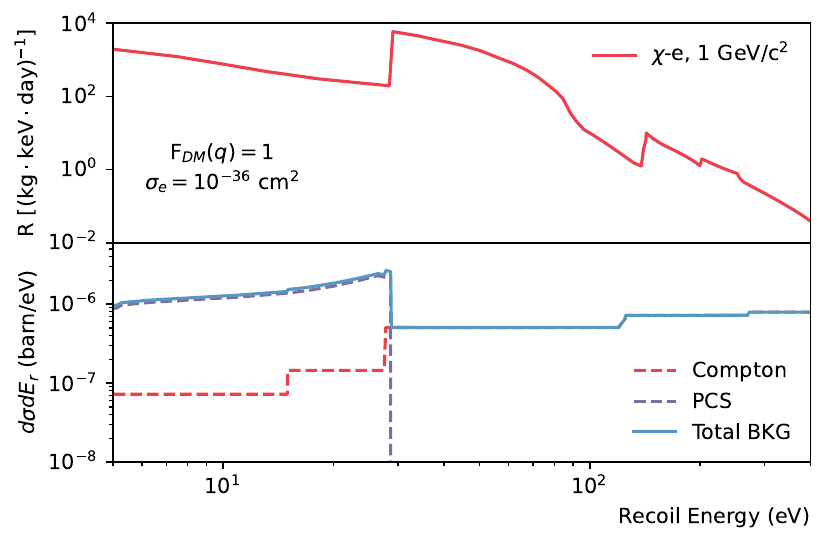}
    \label{fig:chi-e}
  }
  \vspace{-15pt}
  \subfigure[] {
    \includegraphics[width=\linewidth]{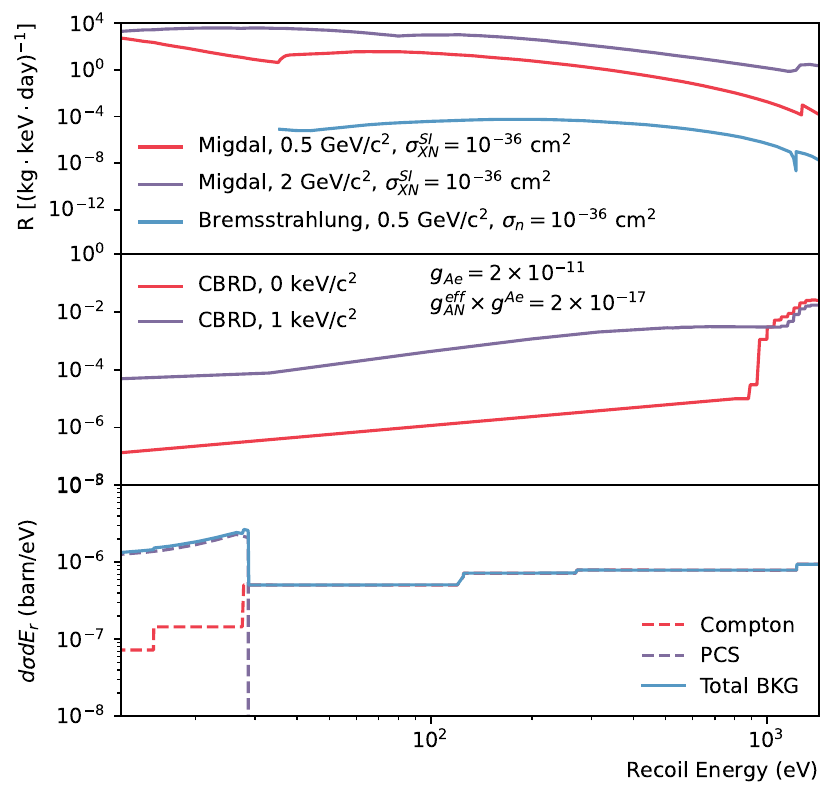}
    \label{fig:cbrd}
  }
  \caption{The typical LDM expected spectra via electronic final state and
    $\gamma$ induced background on HPGe detector.
    Figures~\ref{fig:alp-dp}, \ref{fig:chi-e} and \ref{fig:cbrd} represent
    for different interested energy regions for corresponding candidates.}
\end{figure}

Mass differences result in the most significant variations in spectral
shapes.
These variations primarily arise from multiple CS and
electron escape effects.
In a 5 g HPGe detector, the Compton background is predominantly
influenced by single scattering, producing a spectrum that is highly
correlated with DCS and displaying a step slope below incident photon
energy of approximately 400 keV~\cite{qiaoComptonScatteringEnergy2020}.
Conversely, in a 1 kg HPGe detector, energy deposition is more
likely to concentrate in the high-energy region due to multiple scattering
and enhanced photoelectric effects, which significantly reduce the
slope of Compton spectra.
The step slope vanishes for incident photons exceeding 400 keV.
Photons from $^{40}\mathrm{K}$ and $^{208}\mathrm{Tl}$ transfer more energy
to the electron system, enabling electrons to transport considerable
distances and escape from the HPGe detector.
This escape phenomenon is more pronounced in the 5 g detector, resulting
in a non-flat background spectrum.
As illustrated in Fig.~\ref{fig:compt-bkg} (middle), more than half of
the low-energy events arise from incomplete energy deposition due to
electron escape in 5g HPGe detector.
Small-angle scattering events with complete energy deposition are
expected to contribute a flat background, consistent with the predictions
of the differential cross-section (DCS).
For 1 kg detector, the non-flat structure caused by electron escape
diminishes.

As illustrated in Fig.~\ref{fig:compt-bkg} (bottom), the structure
differences on spectra raised by source positions are minor
(few percent) on the energy spectra and
shows no significance dependence on incident energies.

\subsection{Background for LDM Electronic Recoil Channels}
\label{sec:channels-bkgs}
We performed an combined $\gamma$ induced electronic final-state background
analysis for varies LDM models.
DM models raised from different theoretical motivations yield distinct
expected energy spectra in HPGe detector.
Their unique spectral features can be distinguished from the background,
particularly at binding energies.
The LDM search from the electronic final states covers spanning from
GeV/$c^2$ to MeV/$c^2$.
The LDM candidates are categorized according to their expected energy
spectra and corresponding background structures.
The ionization quenching of PCS is considered via Lindhard
model~\cite{lindhard1963integral}, where the paremeter $k=0.162$ is
determined through recent
measurement~\cite{bonhommeDirectMeasurementIonization2022}.
The background uses 2614 keV photon energy coherent and CS
spectrum.

Fig.~\ref{fig:alp-dp} illustrates that the most prominent structure in
expected spectra of ALPs~\cite{wangImprovedLimitsSolar2020} and dark
photons~\cite{sheDirectDetectionConstraints2020} exhibit around
the L-shell ionization energy, where only CS contributes to the
background.
The cross-sections of both ALPs and dark photons display abrupt
variations at the ionization energies of atomic shells, as their
scattering processes are analogue to the photoelectric effect.
Although these two DM candidates exhibit similar structures at the L-shell
ionization edge of germanium, the steps in the expected spectra are
distinguishable from background, as the steps of signals are more
pronounced than the Compton steps.

The $\chi$-e spectrum (Fig.~\ref{fig:chi-e}) considered four types of
electron transition, which leads to several discontinuities at about
29 eV, 140 eV, 200 eV and 252 eV~\cite{zhangConstraintsSubGeVDark2022}.
The PCS cutoff and Compton step of Ge $M_\mathrm{IIIa(b)}$ shell are near
29 eV, which lead to a 6$\times$ step height in total.
However, the step height of $\chi$-e is more pronounced
(approximately ${30\times}$).
At the remaining spectral features in the $\chi$-e spectrum, $\gamma$-induced
backgrounds provide a flat continuum.

Fig.~\ref{fig:cbrd} shows the spectrums of CBRD axion and $\chi$-N scattering
via Migdal effect~\cite{liuConstraintsSpinIndependentNucleus2019,
  ibeMigdalEffectDark2018}.
Both of the CBRD spectrums ascend quickly near 1 keV, while structure of
Compton background is relative flat.
The expected Migdal effect and Bremsstrahlung spectrum have discontinuity
at 1.2 keV corresponding to L-shell of Ge.
The expected signals for ALPs and CBRDs span a very broad energy range.
The expected signals for ALPs and CBRDs span a broad energy range.
However, the slowly varying expected signals in the tens of eV region face
a rapidly rising PCS background, making discoveries in this energy range
particularly challenging.

\section{Summary}

This work investigates the $\gamma$-induced backgrounds for direct dark matter
detection involving an electronic final states, taking into account the
effects of CS and PCS.
The aim is to provide a robust understanding of the backgrounds for
current and next-generation HPGe experiments.

CS dominates the background in the energy range from
sub-keV to the K-shell ionization energy.
A fully relativistic, atomic many-body ab initio calculation is performed
to reassess CS, revealing at most a factor of two
difference compared to the previous
study~\cite{hubbellAtomicFormFactors1975}.
To clarify the low-energy CS behavior, we designed
an experiment using HPGe and NaI[Tl] detectors to precisely measure the
DDCS at six scattering angles from $1.5^\circ$ to $12^\circ$.
Accurate calibration of the geometric angles ensures that the errors are
controlled within $0.03^\circ$.
The efficiencies, background, and systematic errors in the low-energy
region are meticulously considered.

The measurements provided the capability to elucidate the inconsistencies
among three low-energy Compton models implemented in \geant: the
Livermore~\cite{apostolakisGEANT4LowEnergy1999},
Penelope~\cite{salvatgavaldaPENELOPE2008Code2009}, and
Monash~\cite{brownLowEnergyBound2014} models.
For measurements across all scattering angles, the experimental data
evidently reject the Monash model (beyond $5\sigma$ significance), except for
the, except for the measurement at $1.5^\circ$ ($4.4\sigma$).
The lower significance of the $1.5^\circ$ data  is attributed to the
suppression of the most significant differences by the low-energy PSD
efficiency.
The Livermore model (and the Penelope model) demonstrates consistency with
the measurements; however, a mild overestimation is observed in the
low-energy region ($<$ 500 eV).
This discrepancy may arise from the assumption of isolated atoms.
In pracitce, the outer shell electrons should be considered using energy
band formalism, which leads to differences in the electronic structure.
Measurements on the SF are insufficient to clarify the discrepancy but
favor the MCDF-RIA result.
Investigating the lower momentum transfer region in SF necessitates a
well-collimated source to minimize events with large scattering angles,
as well as a low-threshold HPGe detector to capture a greater number of
low-energy events.

The impacts of SF discrepancies, detector mass as well as $\gamma$ source
position on DM background are evaluated through \geant simulation.
The size of detector significantly influence the shapes of Compton
spectra, leading to non-flat background structures due to non-negligible
electron escaping at high $\gamma$ energies.
Furthermore, as anticipated, the MCDF-RIA SF predicts a more pronounced
background level, about 10\% to 50\% higher than the results from HF-RIA
relatively, primarily due to its preference for small-angle scattering.
However, no significant differences are observed concerning the emitting
position of $\gamma$-ray.

The analysis of the combined $\gamma$-induced background against LDM
candidates, which includes CS and PCS, has been conducted.
Although similar structures appear at sub-shell ionization energies for
both expected signals and backgrounds, the $\gamma$-induced background has a
limited likelihood of being misidentified as signals.

\begin{acknowledgments}
  This work was supported by the National Key Research and Development
  Program of China (Contract No. 2023YFA1607103) and the National Natural
  Science Foundation of China (Contracts No. 12441512, No. 11975159,
  No. 11975162) provided support for this work.
\end{acknowledgments}

\bibliography{ref.bib}

\end{document}